\documentclass[twocolumn]{aastex63}
\usepackage{newtxtext,newtxmath}
\usepackage[T1]{fontenc}
\usepackage{CJKutf8}
\usepackage{ae,aecompl}
\usepackage{graphicx}
\usepackage{amsmath}
\usepackage{amssymb}	
\usepackage{float}
\usepackage[]{hyperref}

\usepackage{booktabs}
\usepackage{float}

\accepted{in ApJ}

\begin{document} 

\title{On the robustness of the velocity anisotropy parameter in probing the stellar kinematics in Milky Way like galaxies: Take away from TNG50 simulation}

\shorttitle{Velocity Anisotropy}
\shortauthors{R. Emami et. al.}

\correspondingauthor{Razieh Emami}
\email{razieh.emami$_{-}$meibody@cfa.harvard.edu}

\author[0000-0002-2791-5011]{Razieh Emami}
\affiliation{Center for Astrophysics $\vert$ Harvard \& Smithsonian, 60 Garden Street, Cambridge, MA 02138, USA}

\author[0000-0001-6950-1629]{Lars Hernquist}
\affiliation{Center for Astrophysics $\vert$ Harvard \& Smithsonian, 60 Garden Street, Cambridge, MA 02138, USA}

\author[0000-0001-8593-7692]{Mark Vogelsberger}
\affiliation{Department of Physics, Kavli Institute for Astrophysics and Space Research, Massachusetts Institute of Technology, Cambridge, MA 02139, USA}

\author[0000-0002-6196-823X]{Xuejian Shen}
\affiliation{TAPIR, California Institute of Technology, Pasadena, CA 91125, USA}

\author[0000-0003-2573-9832]{Joshua S. Speagle (\begin{CJK*}{UTF8}{gbsn}沈佳士\ignorespacesafterend\end{CJK*})}
\altaffiliation{Banting \& Dunlap Fellow}
\affiliation{David A. Dunlap Department of Astronomy \& Astrophysics, University of Toronto, 50 St. George Street, Toronto ON M5S 3H4, Canada}
\affiliation{Dunlap Institute for Astronomy and Astrophysics, University of Toronto, 50 St George Street, Toronto, ON M5S 3H4, Canada}
\affiliation{Department of Statistical Sciences, University of Toronto, 100 St George St, Toronto, ON M5S 3G3, Canada}

\author[0000-0002-3430-3232]{Jorge Moreno}
\affiliation{Department of Physics and Astronomy, Pomona College, Claremont, CA 91711, USA}

\author[0000-0002-7892-3636]{Charles Alcock}
\affiliation{Center for Astrophysics $\vert$ Harvard \& Smithsonian, 60 Garden Street, Cambridge, MA 02138, USA}

\author[0000-0002-3185-1540]{Shy Genel}
\affiliation{Center for Computational Astrophysics, Flatiron Institute, New York, USA}
\affiliation{Columbia Astrophysics Laboratory, Columbia University, 550 West 120th Street, New York, NY 10027, USA}

\author[0000-0002-1975-4449]{John C. Forbes}
\affiliation{Center for Computational Astrophysics, Flatiron Institute, New York, USA}

\author[0000-0003-3816-7028]{Federico Marinacci}
\affiliation{Department of Physics \& Astronomy "Augusto Righi", University of Bologna, via Gobetti 93/2, 40129 Bologna, Italy}

\author[0000-0002-5653-0786]{Paul Torrey}
\affiliation{Department of Astronomy, University of Florida, 211 Bryant Space Sciences Center, Gainesville, FL 32611, USA}

\begin{abstract}
We analyze the velocity anisotropy of stars in real and energy space for a sample of Milky Way-like galaxies in the TNG50 simulation. We employ different selection criteria, including spatial, kinematic and metallicity cuts, and make three halo classes ($\mathcal{A}$-$\mathcal{C}$) which show mild-to-strong sensitivity to different selections. The above classes cover 48\%, 16\% and 36\% of halos, respectively. We analyze the $\beta$ radial profiles and divide them into either monotonically increasing radial profiles or ones with peaks and troughs. We demonstrate that halos with monotonically increasing $\beta$ profiles are mostly from class $\mathcal{A}$, whilst those with peaks/troughs are part of classes $\mathcal{B}$-$\mathcal{C}$. This means that care must be taken as the observationally reported peaks/troughs might be a consequence of different selection criteria. We infer the anisotropy parameter $\beta$ energy space and compare that against the $\beta$ radial profile. It is seen than 65\% of halos with very mild sensitivity to different selections in real space, are those for which the $\beta$ radial and energy profiles are closely related. Consequently, we propose that comparing the $\beta$ radial and energy profiles might be a novel way to examine the sensitivity to different selection criteria and thus examining the robustness of the anisotropy parameter in tracing stellar kinematics. We compare simulated $\beta$ radial profiles against various observations and demonstrate that, in most cases, the model diversity is comparable with the error bars from different observations, meaning that the TNG50 models are in good overall agreement with observations. 

\end{abstract}


\section{Introduction}
Halo stars comprise only $\sim1\%$ of the stellar mass in the Milky Way (MW). Nevertheless, they provide remarkable information about the assembly history of our Galaxy. Stars belonging to the stellar halo are typically old and metal-poor, pointing to the halo's ancient origin. By tracking the current orbits of stars, we may get some information on their motions at the early times. Consequently, we may use their current orbit to track the kinematics of their progenitors, such as external satellite galaxies or the gas clouds where these stars formed in initially \citep{2013ApJ...763L..17H}. Furthermore, stellar kinematics can also be employed for studying the distribution of the gaseous component
\citep{1994MNRAS.270..325C, 2005MNRAS.357.1113K,2012ApJ...745...92A, 2014ApJ...789...63A}.

A widely used technique to model the galactic mass distribution is the Jeans dynamical modelling \citep{1915MNRAS..76...70J, 1980MNRAS.190..873B, 1985AJ.....90.1027M, 1992ApJ...391..531D, 2014RvMP...86...47C}. Motivated by this, it is customary to compute the root-mean-square (rms) of the radial velocity as $v_{\rm rms} \equiv \sqrt{\langle v^2_r \rangle}$ \citep{2017ApJ...835..193E} and the velocity anisotropy parameter,
\begin{equation}
\label{beta}
\beta \equiv 1 - \left( \frac{ \sigma^2_{\theta} + \sigma^2_{\phi} }{2 \sigma^2_{r} } \right).
\end{equation}
Here $\sigma_i$ refers to the velocity dispersion along the following directions: $i = (r, \theta, \phi)$, and it is calculated as $\sigma_i \equiv \sqrt{ \langle v^2_i \rangle - \langle v_i \rangle ^2}$. The velocity anisotropy parameter, hereafter $\beta$, was initially introduced by \cite{1980MNRAS.190..873B} to measure the orbital structure of a given system. In this parametrization, $\beta$ ranges from $-\infty$ (for purely tangential orbits) through $0$ (for completely isotropic motions) and to $1$ (for purely radial orbits). Radial orbits correspond to $\beta>0$, whilst tangential ones have $\beta <0$. The velocity anisotropy is vital in accurately measuring the mass distribution in dispersion-supported systems, from dwarf spheroidal galaxies \citep{2007ApJ...663..948G, 2007ApJ...669..676S,2010MNRAS.406.1220W} to massive ellipticals \citep{2005Natur.437..707D}.

Direct measurements of $\beta$ have been somewhat challenging because they require a full three-dimensional map of stellar velocities, which had not been possible until very recently, thanks to the development of Hubble Space Telescope (HST) and the Gaia mission. Observers commonly measure the line-of-sight (LOS) velocity and use this to infer the $\beta$ \cite[see for example][and references therein]{2004AJ....127..914S,2012ApJ...761...98K, 2012MNRAS.424L..44D,2015ApJ...813...89K}. The main reason for this is that owing to our position in the MW galaxy, line-of-sight (LOS) velocities provide useful information about tangential velocities. However, \cite{2017ApJ...841...91H} pointed out that since stellar orbits become progressively more radial at larger galactocentric distances, using only LOS velocities may underestimate the $\beta$. 
Consequently, the velocity anisotropy parameter has instead only been measured indirectly via dynamical modelling \cite[see for example][and references therein]{1994MNRAS.270..325C, 2014MNRAS.443..598D, 2014MNRAS.443..610D}. 

Fortunately, it is now possible to measure tangential velocities for a small sample of stars away from the galactic center by using full three-dimensional velocity measurements \cite[e.g.][and references therein]{2016ApJ...820...18C, 2009MNRAS.399.1223S, 2010ApJ...716....1B, 2012ApJ...753....7S, 2019ApJ...873..118W, 2018A&A...616A..12G}, although getting very accurate LOS velocity may require obtaining the spectra.

In observational studies, authors have adopted different assumptions for the $\beta(r)$ profile. For example, it is commonly assumed that either $\beta(r)$= Const \citep[e.g.,][and references theirin]{2005MNRAS.363..918L, 2007AJ....134..566K,2007ApJ...667L..53W,2008ApJ...681L..13B, 2009MNRAS.394L.102L, 2009ApJ...704.1274W}, or that  
$\beta(r)$ varies radially \citep[e.g.][]{2001ApJ...563L.115K,2004ApJ...611L..21W, 2005MNRAS.363..705M, 2007ApJ...663..948G, 2008ApJ...681L..13B,2013MNRAS.429.3079M,2015arXiv150408273M}. Whilst a constant $\beta$ is well suited to measure the mass of the MW, a radially-varying $\beta$, may have additional power in probing the MW's accretion history \cite[e.g.][]{2019ApJ...879..120C, 2018ApJ...853..196L, 2020arXiv200505980B}. \cite{1990AJ....100.1191M}
used the G and K giants (located up to few kpc from the Sun) and estimated $\beta \approx 0.5$. \cite{1998AJ....115..168C} estimate $\beta = 0.52 \pm 0.07$ from a sample of metal-poor halo stars near the Sun. Beyond the Solar neighborhood, one may also estimate $\beta$ with line-of-sight velocities, by adopting mass-distribution models of the galaxy. Appropriate tracers include blue horizontal branch (BHB) stars, K giants and F-type stars \citep[see e.g.][and references therein]{1994MNRAS.271...94S, 2005MNRAS.360..354T, 2012ApJ...761...98K, 2015ApJ...813...89K}. \cite{2009MNRAS.399.1223S} used halo sub-dwarfs from the Sloan Digital Sky Survey (SDSS) and estimated $\beta = 0.69 \pm 0.01$. \cite{2019AJ....157..104B} used K giants from the Large Sky Area Multi-Object Fiber Spectroscopic Telescope (LAMOST) catalog and measure $\beta$ as a function of Galactocentric radius. 

On the theory front, simulations of MW-like galaxies often result in radially-increasing $\beta$ profiles  \cite[e.g.][]{2005MNRAS.364..367D, 2006MNRAS.365..747A,2007MNRAS.379.1464S, 2013ApJ...773L..32R, 2012ApJ...761...98K, 2013MNRAS.428..129S}. 
As shown by \cite{2018ApJ...853..196L}, analyses of this kind unveil more complex structures in $\beta$ and in the halo in general \cite[e.g.][]{2012ApJ...746...34B, 2012A&A...538A..21S, 2017ApJ...841...59Z}.

In this paper, we analyze the stellar kinematics inferred from the stellar velocity anisotropy parameter, $\beta$, in a sample of Milky Way (MW) like galaxies from the TNG50 run of the Illustris-The-Next-Generation (\texttt{Illustris-TNG}) project. The enhanced resolution in TNG50 makes it possible to analyze radial $\beta$ profiles very accurately, near the level of the zoom-in simulations and also consider the impact of cosmic evolution and mergers of a larger statistical population as TNG50 simulation has many more galaxies in the box. 
We analyze the $\beta$ radial and energy profiles at redshift $z=0$ and investigate their dependencies on various selection criteria, including spatial and kinematic cuts as well as the ones coming from some metallicity cuts. 

We design a halo classification based on the sensitivity levels of $\beta$ profile to various selection criteria and place halos in three classes. $\mathcal{A}$ class refers to halos with very mild sensitivity to different selections. Halos part of class $\mathcal{B}$ show some levels of sensitivities to different selections and those associated with $\mathcal{C}$ show a very strong dependencies to model selections. We show that 48\%, 16\% and 36\% of halos are part of classes $\mathcal{A}$ to $\mathcal{C}$, respectively. We analyze the $\beta$ radial profile and show that it can be divided to monotonically increasing profiles or the ones with peaks/troughs, which contain 32\% and 68\% of the halos, respectively. It is shown that, the $\beta$ profile from the monotonically increasing class smoothly increases from the interior to the exterior part of the halo. While, the $\beta$ from the class with peaks/troughs experiences some local fluctuations. We further demonstrate that almost all of the halos from the first class are part of category $\mathcal{A}$, with only one exception. On the contrary, members of the class with peaks/troughs are mostly part of category $\mathcal{B}$ and $\mathcal{C}$, with only 29\% of them being part of category $\mathcal{A}$. This means that care should be taken in 
interpreting the observed peaks/troughs as they might depend on the actual selection criteria. 

We infer the impact of different metallicity and eccentricity cuts in the $\beta$ radial profile and also investigate the impact of different stellar orbital types such as the radial, prograde and retrograde on the $\beta$ radial profile. 

Moving in the energy space, we infer and compare the $\beta$ radial and energy profiles against each other and make another halo classification based on their differences. It is seen that 65\% of the halos with very mild sensitivity on different selection criteria, in the real space, are among those for which the $\beta$ radial and energy profiles are closely related. On the contrary, halos in which the $\beta$ radial and energy profile are rather different are entirely coming from category $\mathcal{C}$. This again establishes that one should be careful when drawing any conclusions from these classes. Furthermore, we propose that one quantitative way to check if the observational results are very robust is to also compute the $\beta$ energy profile and compare that with the $\beta$ radial profile. Based on our conclusion it is likely that those with very similar profiles are also very robust against altering the selection criteria. 

It must be emphasized that our current $\beta$ analysis in the real and energy spaces, and their correlations, are purely theoretical and are based on a full awareness of the gravitational potential for the MW. From the observational perspective, we may only estimate the potential to some extent. Furthermore, the current paper is not aimed to prove that we identify an unbiased sample of stars using this comparison. But rather that for a given sample of stars such a comparison between the $\beta$ radial and energy profiles likely indicates how the sample may show sensitivities to changing different selection criteria.

Motivated by the aforementioned halo classification, throughout this paper, we make a sub-halo sample of 4 galaxies from each of the above classes and study their behavior thoroughly. This includes halos [4, 12, 20, 25] which are the members of Category [$\mathcal{A}$, $\mathcal{B}$, $\mathcal{B}$, $\mathcal{C}$], respectively. 

Ultimately, we overlay our theoretical outcomes, with few different selection criteria, on top of the most recent observational results and compare them with each other. Our comparison is made both at the level of isolated data points as well as the extended observations in which the $\beta$ profile is reported in more than one single point. It is shown that there are good levels of agreement between the two both for the isolated points and for the extended observations where we see for each halo there are some data points that easily pass through the profile. 

This paper is organized as follows. Section~\ref{TNG} summarizes the TNG50 simulation and our sample of MW-like galaxies. Section~\ref{star-metal} links stellar metallicity to other stellar properties. Section~\ref{beta-prof}
computes the radial and energy velocity anisotropy profiles. Section~\ref{observations} compares our simulated radial-$\beta$ profiles against observations. Lastly, Section~\ref{concl} presents our conclusions. 

\section{TNG50 simulation}
\label{TNG}
TNG50 is the highest resolution run of the suite of large-scale series of IllustrisTNG simulations \citep[e.g.][]{2019MNRAS.490.3196P, 2019MNRAS.490.3234N}. It provides an excellent combination of the volume and resolution close to the level of zoom-in simulations \citep[e.g.,][]{2016MNRAS.459L..46M, 2018MNRAS.481.1726G, 2019MNRAS.488..135H} and thus provides a natural avenue to investigate the impact of galaxy evolution with very high resolution. It evolves supermassive black holes (SMBHs), dark matter, gas, stars and magnetic fields within a periodic boundary volume of 51.7 kpc$^3$. Its softening length is 0.39 comoving kpc $h^{-1}$ for $z\geq1$, reducing to 0.195 proper kpc $h^{-1}$ for $z<1$. 

In TNG50, the model for galaxy formation 
includes a stochastic, gas-density-threshold-based for the star formation, the evolution of mono-age stellar populations which are being represented by the star particles, the chemical enrichment of the Inter Stellar Medium (ISM) as well as the tracking of 9 different chemical elements including (H, He, C, N, O, Ne, Mg, Si, Fe) in addition to total gas metallicity and the Europium. It also includes the gas cooling and heating, the feedback from supernovae which is in the form of galactic winds, seeding and the growth of supermassive black holes (SMBHs) and the injection of the energy as well as the momentum from SMBHs into the surrounding gas.

Using various selection criteria, such as dark matter halo mass and choosing the rotationally-supported stellar population in \citep{2020arXiv201212284E,2021ApJ...913...36E, 2021AAS...23832407E}, we created a sample of 25 MW-like galaxies in the TNG50 simulation. This paper probes the structure of the stellar distribution, ignoring the impact of satellites, with stellar kinematics as the tracer. We also analyze different stellar properties, including stellar age, metallicty and the stellar velocity anisotropy profile. One of our goals is to investigate how different selection criteria based on these quantities impact radial velocity anisotropy profiles. 

\begin{table*}[!htbp] \centering
\caption{Linking between the subhalo ID with the galaxy number in our sample of MW like galaxies in TNG50.}
\label{halo_ID} 
\begin{tabular}{|l|c|c|c|c|c|c|c|c|r|} 
\hline 
$1 \mapsto 476266$ & $ 2 \mapsto 478216$ & $ 3 \mapsto 479938$ &   $4 \mapsto 480802$ &  $5 \mapsto 485056 $  \\
\hline 
$6 \mapsto 488530$ & $7 \mapsto 494709 $ & $8 \mapsto 497557$ & 
$9 \mapsto  501208$ & $10 \mapsto 501725 $ \\
\hline 
$ 11 \mapsto 502995 $ & $ 12 \mapsto 503437$ &   $13 \mapsto 505586$ &  $14 \mapsto 506720 $ & $15 \mapsto 509091$ \\
\hline 
$16 \mapsto 510585 $  &
 $17 \mapsto 511303$ & $18 \mapsto  513845$ & $19 \mapsto  519311 $ & $ 20 \mapsto  522983 $ \\
\hline 
 $ 21 \mapsto 523889 $ & $ 22 \mapsto  529365 $ & $ 23 \mapsto  530330$ & $ 24 \mapsto  535410 $ &
 $ 25 \mapsto  538905 $  \\
 \hline 
\end{tabular}
\end{table*}

\begin{figure}
\center
\includegraphics[width=0.45\textwidth]{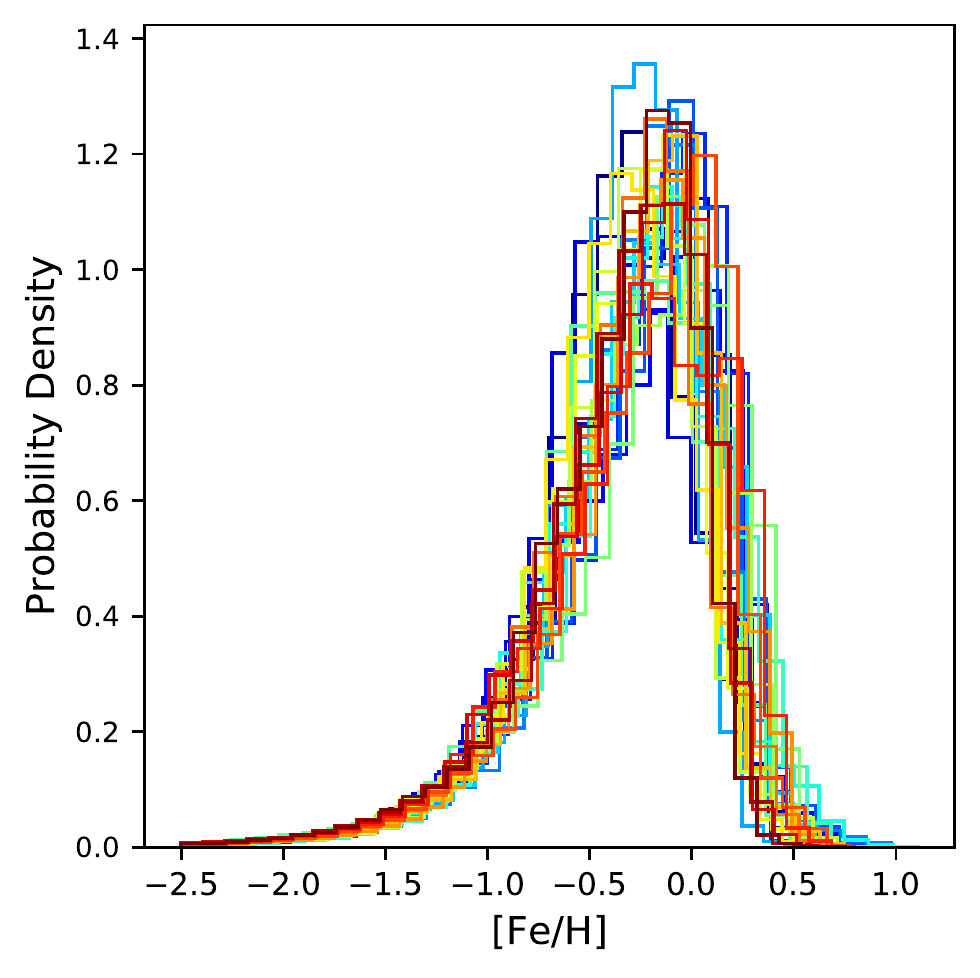}
\caption{The probability density distribution of stellar metallicity in our MW-like galaxy sample. It is evident that stars from different galaxies have quite similar metallicity distribution. Different colors refer to various galaxies. Low metallicities are in the tail of the distribution.}
\label{Fe-dist}
\end{figure}

\section{Stellar Metallicity} 
\label{star-metal}
Since the metallicity cut is one of the key ingredients in our stellar sample selection, in what follows we study the correlation between stellar metallicity and various other stellar properties, including the stellar age, the spatial metallicity distribution and the stellar velocity. 

First, we present the stellar metallicity distribution. Figure \ref{Fe-dist} presents the 1D probability density distribution of all of stars in different galaxies in our sample. Quite interestingly, stars from different galaxies in our sample seem to have very similar metal distribution. Moreover, it is evident that stars with low metallicty are in the tail of the distribution, whilst those with [Fe/H] in the range [-0.5-0.0], sit on the peak of the distribution. It is intriguing to see how different metallicty-based cuts might affect the velocity anisotropy of stars. In Sec. \ref{beta-metal}, we get back to this question in depth. 

\begin{figure*}
\center
\includegraphics[width=1.0\textwidth]{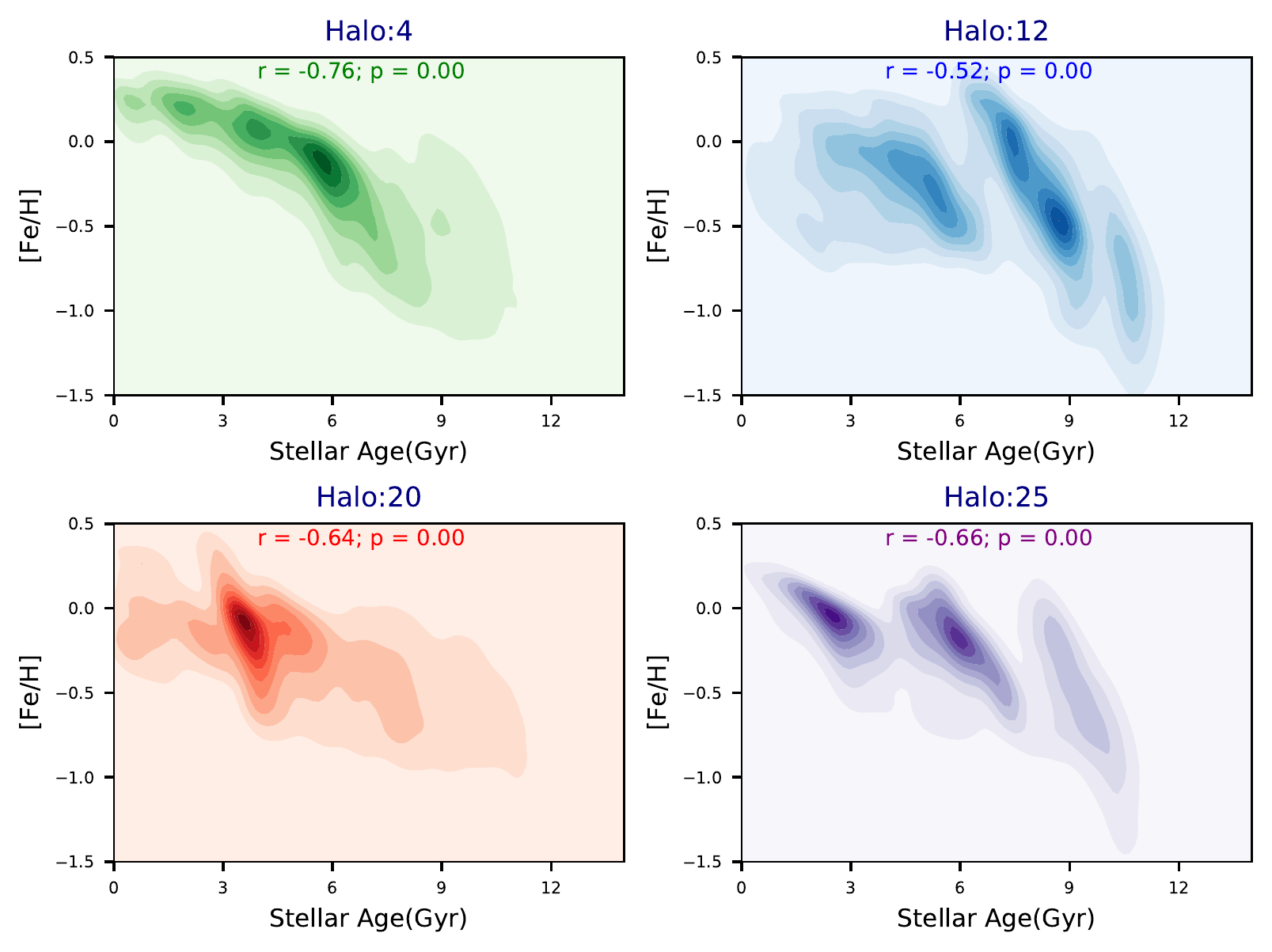}
\caption{2D Correlation of stellar age vs metallicity for a sub-sample of four galaxies from our MW-like galaxy sample. The Spearman correlation is negative indicating that metal-poor stars are older than the metal-rich ones.}
\label{Fe-Age}
\end{figure*}

\subsection{Stellar age vs metallicity}
\label{age-metal}
Next, we study the correlation between the stellar age and the stellar metallicity. Figure \ref{Fe-Age} presents $\mathrm{[Fe/H]}$ versus stellar age for a sub-sample of four galaxies from our sample, including galaxies [4, 12, 20, 25]. 
As already stated above, these halos are members of distinct categories. In each case, we have inferred the Spearman correlation, hereafter r, between the stellar metallicity and the age. 

From the plot, it is inferred that while in some cases, such as galaxy 4, the stellar metallicity is smoothly declining with increasing stellar age, others such as galaxy 12 show a break and thus a discrete profile of age vs [Fe/H] indicating that their associated stars might have different origins. For instance, they might be accreted as a satellite merged onto the galaxy or have arisen from star forming gas out of a galaxy merger etc. Intriguingly, such a break reduces the Spearman correlation as we end up having a bi-modal age-[Fe/H] relation. 

Furthermore, it is seen that in some cases, such as galaxy 20, there is a peak of star formation at an intermediate time, around 4.5 Gyr, that might be an indication of a major merger in the history of the galaxy. Comparing the Spearman correlation of galaxy 4 and 20, we may argue that such a local peak may have slightly diminished the r in galaxy 20. This is better seen when we compare these with galaxy 25, in which we have few distinct bursts of stars and that its |r| is slightly higher than galaxy 20 while it is still less than  the one for galaxy 4. 

In summary, our analysis demonstrate that in all cases, $|r| \geq 0.5$, showing a moderate to large negative correlation between the stellar age vs metallicity depending on galaxy ancient history. Owing to the above correlation between the stellar age vs metallicty, different metallicity cuts naturally pick stars with different ages and allow us to indirectly probe the velocity distribution of young vs old stars. In Sec. \ref{beta-metal}, we address this question. 

\begin{figure*}
\center
\includegraphics[width=0.99\textwidth]{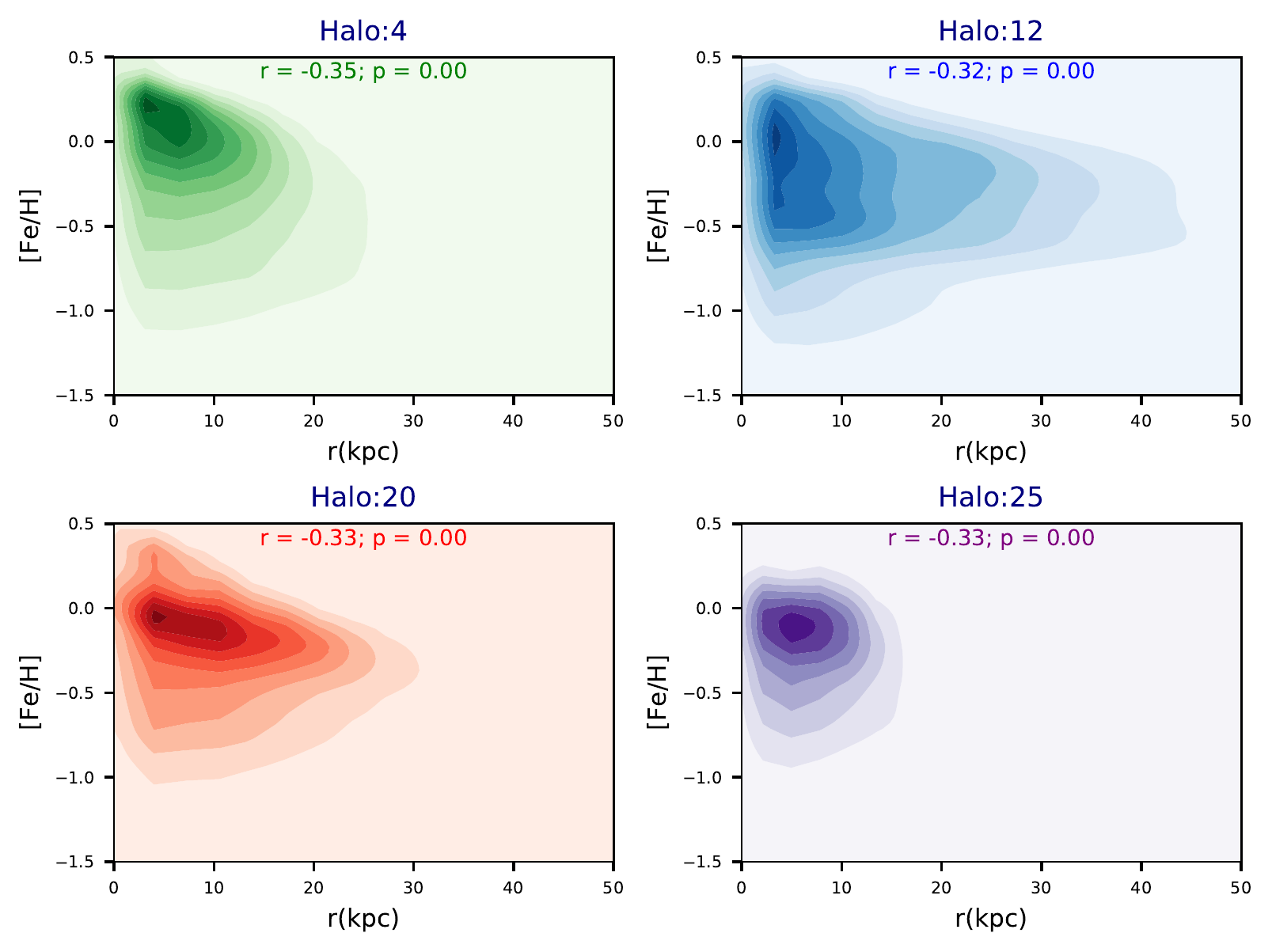}
\caption{2D Correlation of the stellar distance vs stellar metallicity for a sub-sample of four galaxies from our MW-like galaxy sample. The Spearman correlation is modest and negative indicating metal-rich stars are located closer to the galactic center.}
\label{Metal-radius}
\end{figure*}

\subsection{Spatial distribution of metallicity}
\label{radius-metal}
Below we explore the possible correlation between the location of stars and the stellar metallicity. Figure \ref{Metal-radius} presents the 2D distribution of the stellar radii vs the stellar metallicity for a sub-sample of 4 out of our galaxy sample. From the plot, it is inferred that metal-rich stars are generally closer to the galaxy center than the metal-poor ones, in line with the observations \cite[e.g.][]{2015RAA....15.1240H, 2017A&A...600A..70A, 2020ApJ...896...75S, 2020ApJ...894...34D}. Combining that with the above age-metallicity relation, we argue that galaxies in our sample mostly have inside-out star formation in agreement with the recent studies of the inside-out growth of stellar disk \cite[see e.g.][and references therein]{2019ApJ...884...99F, 2021MNRAS.503.1815B, 2021arXiv210309838J}.
The slope of the metallicity-gradient varies from galaxy to galaxy. Some galaxies show a more extended metallicity gradient whilst others have a more concentrated profiles. Furthermore, the extent to which we infer the dominant population of stars is also correlated with their age-metallicity profile.  For instance, galaxy 12 shows a more extended profile inline with our former argument of being affected by mergers. This makes sense as galaxy mergers are expected to expand and dis-locate stars from their original place. Finally, the Spearman correlation between the stellar location vs the stellar metallicity is relatively low, indicating that there are some other factors in locating the stars than their metallicity. 

\begin{figure*}
\center
\includegraphics[width=0.99\textwidth]{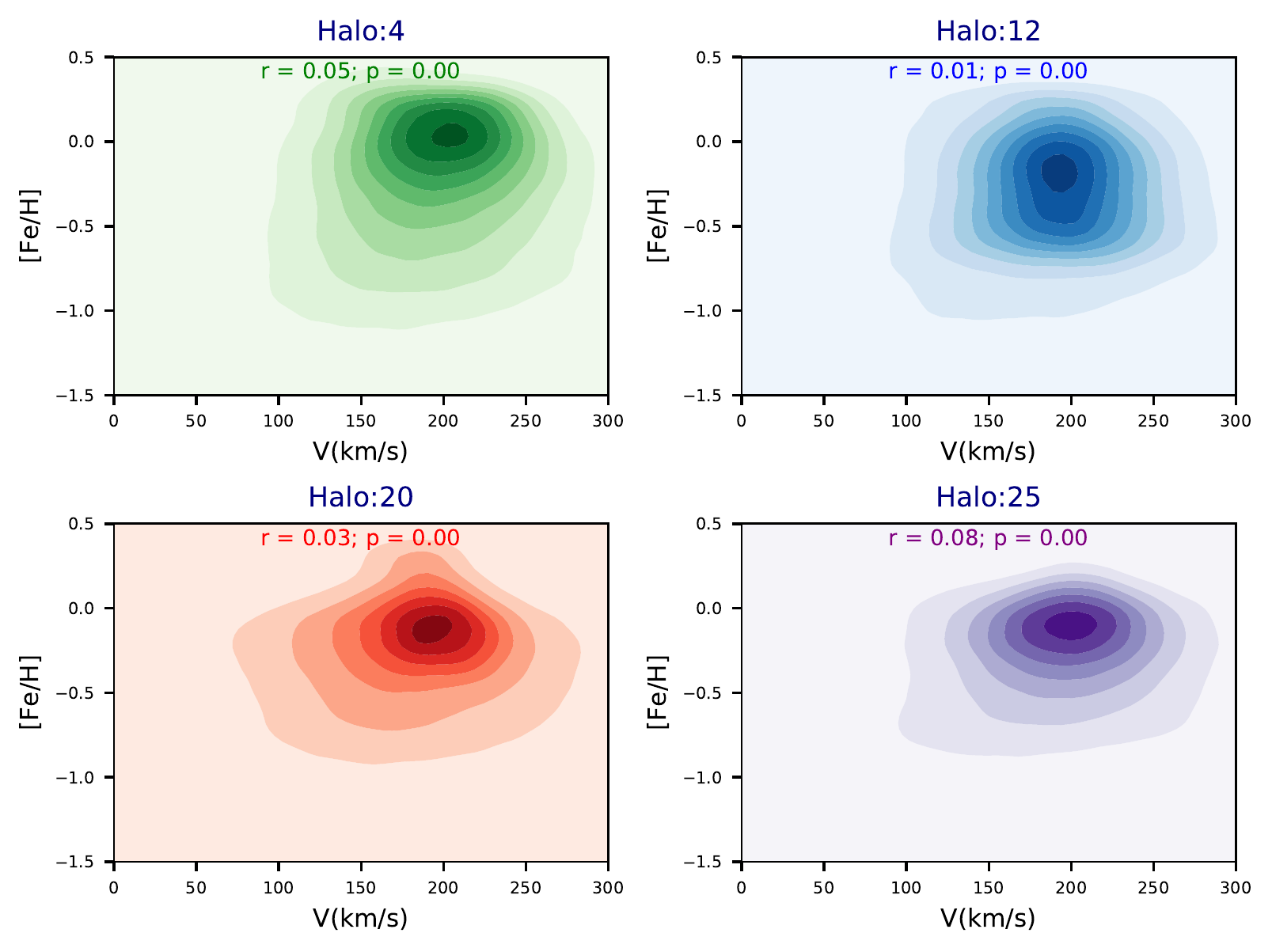}
\caption{2D Correlation of the stellar velocity amplitude vs the stellar metallicity for a sub-sample of four galaxies from our MW-like galaxy sample. It is inferred that the stellar velocity is largely uncorrelated with the stellar metallicity as stellar speed is mainly concentrated between  200-300 km sec$^{-1}$ in agreement with the observed rotational velocity of stars in our Milky Way galaxy.}
 \label{Velocity-Metal}
\end{figure*}

\subsection{Metallicity vs velocity}
\label{metal-v}
As a final step, here we study the correlation between the stellar speed vs the stellar metallicity. Figure \ref{Velocity-Metal} presents the 2D distribution of the stellar velocity amplitude vs the stellar metallicity for a sub-sample 4 out of our galaxy sample. In each case, we infer the Spearman correlation between these quantities demonstrating that they are uncorrelated. 
This makes sense as our galaxy sample obeys the well-known Tully-Fisher relation \citep{1977A&A....54..661T},
as well as the mass-metallicity relation  \citep{2005MNRAS.362...41G, 2008ApJ...681.1183K}. The velocity peaks stand between 200-300 km sec$^{-1}$, indicating that the kinematic of stars inferred in this sample of MW like galaxies are consistent with the rotational velocity of stars seen in  Milky Way disk. 
\begin{figure*}
\center
\includegraphics[width=0.99\textwidth]{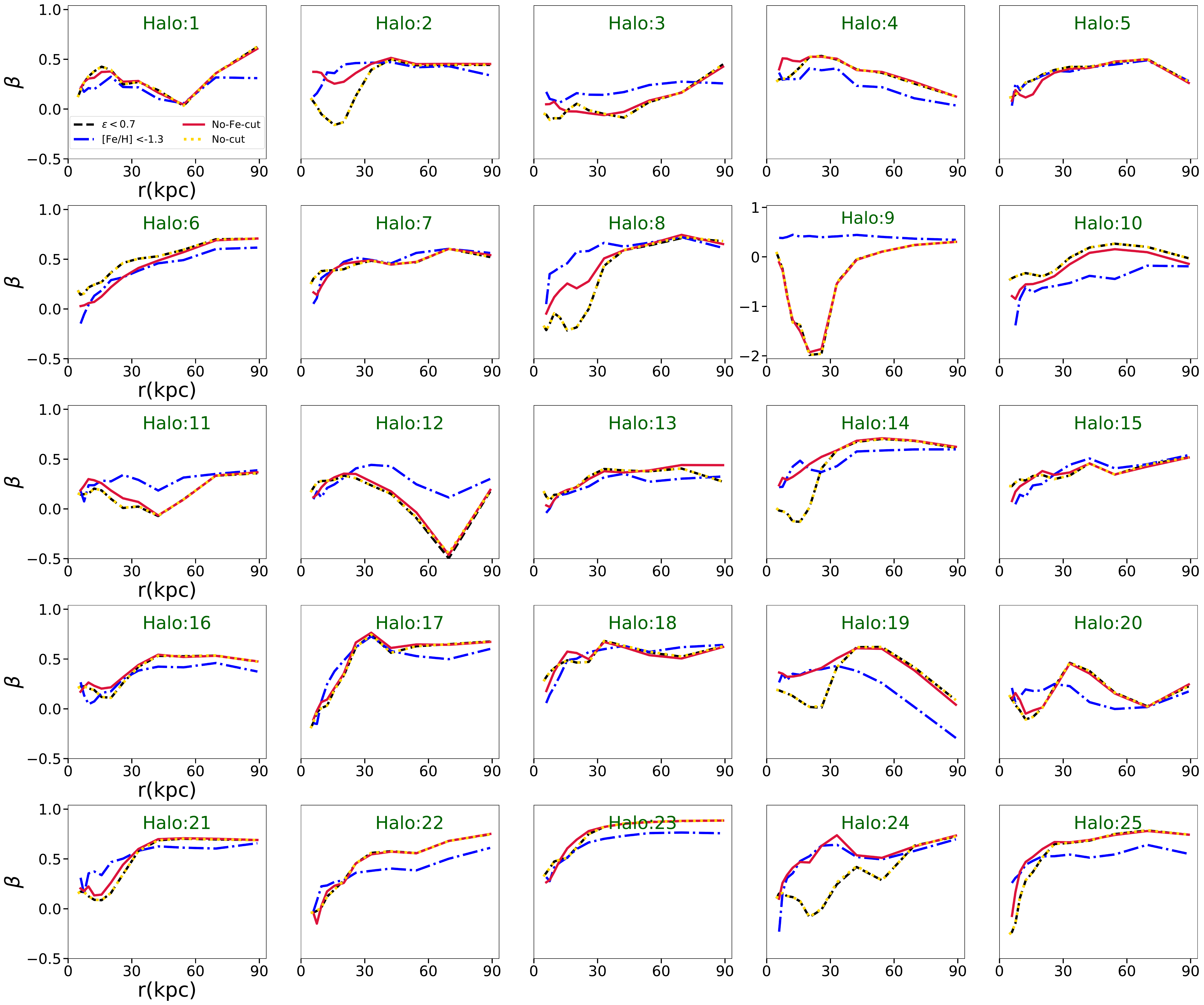}
\caption{$\beta$ radial profile for stars inferred using different selection criteria. Dashed-black, blue-dotted-dashed, solid-red and dotted-yellow lines present $\beta$ inferred from stars with kinematic, metallicity based, spatial cut and no-cut as was used in computing the $\beta$, respectively. }
 \label{beta-radial-profile}
\end{figure*}

\section{Velocity anisotropy: $\beta$} 
\label{beta-prof}
In this section, we analyse the radial profile of the velocity anisotropy as defined in Eq. (\ref{beta}). In our calculations, we use various selection criteria. 
Recall that the $\beta$ parameter characterizes various stellar orbital types. We present the 2D distribution map for the stellar location out of different selection criteria. Furthermore, we study the impact of changing the metallicity cuts, eccentricity and different orbital types on $\beta$ radial profile, $\beta(r)$. 
Finally, we analyse the $\beta$ energy profile and compare that with the $\beta$ radial profile. 

\subsection{The $\beta$ radial profile}
\label{beta-profile}
To infer the $\beta(r)$ for each galaxy, we adopt a coordinate system where the z-axis  (oriented along with the total angular momentum of stars) is orthogonal to the disk, with (x-y) being chosen randomly (right-handed and orthogonal to each other) on the disk plane. As a next step, since $\beta(r)$ deals with  radial and  tangential vectors, we make a transformation to the spherical coordinate system (for both of the coordinate and the velocity vectors). We adopt 15 linearly-distributed radial bins between 1 and 100 \rm{kpc} and, for each of them, we compute the radial and tangential velocity dispersions. We force each bin to contain at least 100 stellar particles (in practice, many of them have over 1000 stellar particles). We choose different selection criteria for stars. Figure \ref{beta-radial-profile} presents the $\beta$ radial profile for 4 different cases as listed below: 

$\bullet$ The dashed-black-line refers to the kinematically-chosen stars with $\epsilon<0.7$, where $\epsilon$ is defined by Eq. (3) in \cite{2021ApJ...913...36E}. Theoretically, this means that we are choosing halo stars \citep[see e.g.][and references therein]{2019MNRAS.485.2589M}. We emphasize that the above threshold is not unique. Nevertheless, it is robust to small variations.

$\bullet$ The blue-dotted-dashed-line describes a hybrid selection of $\mathrm{[Fe/H]}<-1.3$ with $|Z| \geq 5$ \rm{kpc}. This choice is based on the LAMOST data \citep{2019AJ....157..104B}. 

$\bullet$ With the solid-red-lines, we relax the metallicity cut and only choose stars above the disk plane $|Z| \geq 5$ \rm{kpc}.

$\bullet$ The dotted-yellow-line refers to the case with no spatial cut.

At the first glance, it seems that in some cases the velocity anisotropy of stars with completely different selection criteria might be somewhat close. 

\begin{figure*}
\center
\includegraphics[width=0.995\textwidth]{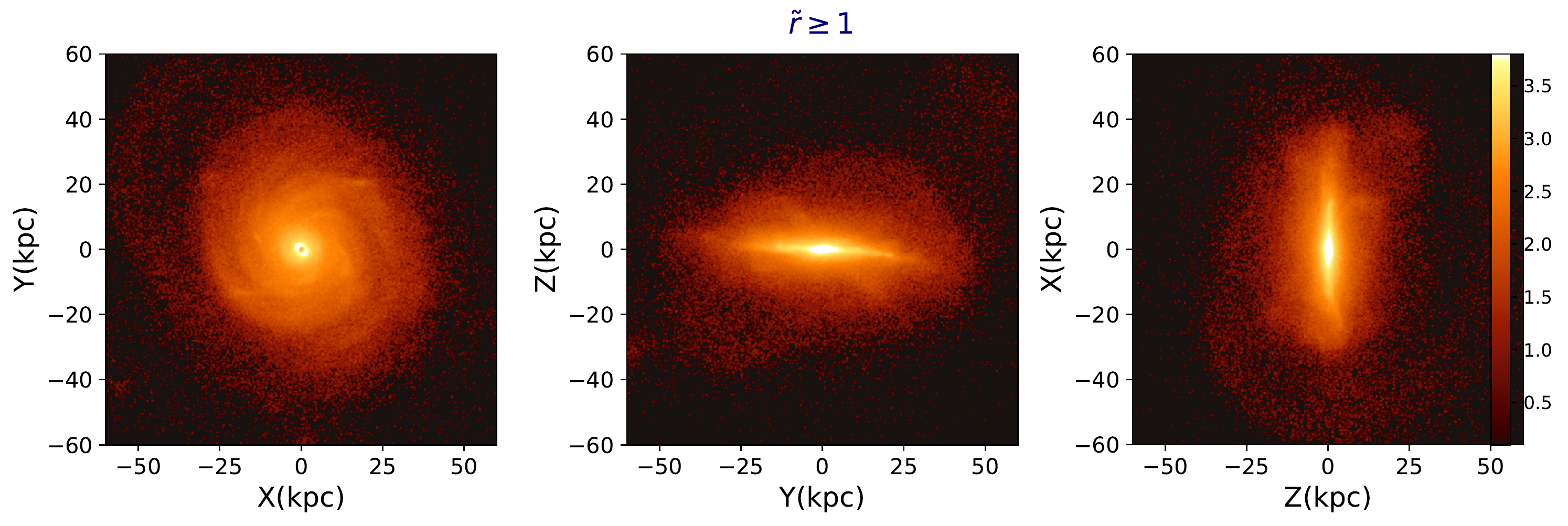}
\includegraphics[width=0.995\textwidth]{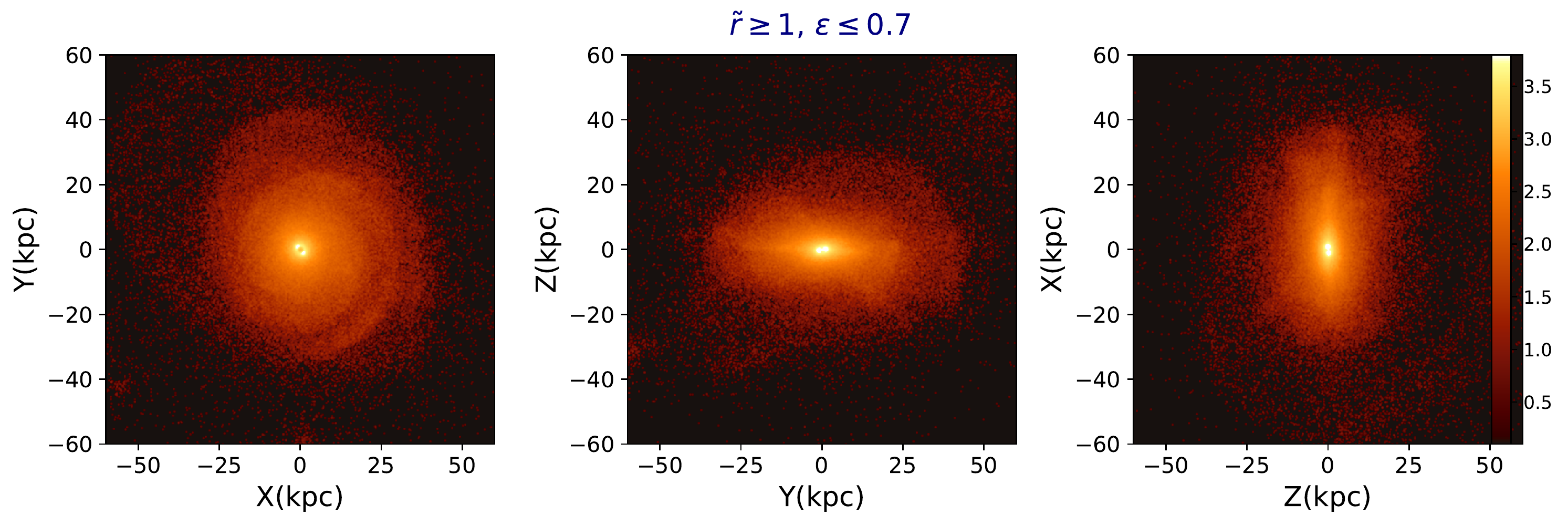}
\includegraphics[width=0.995\textwidth]{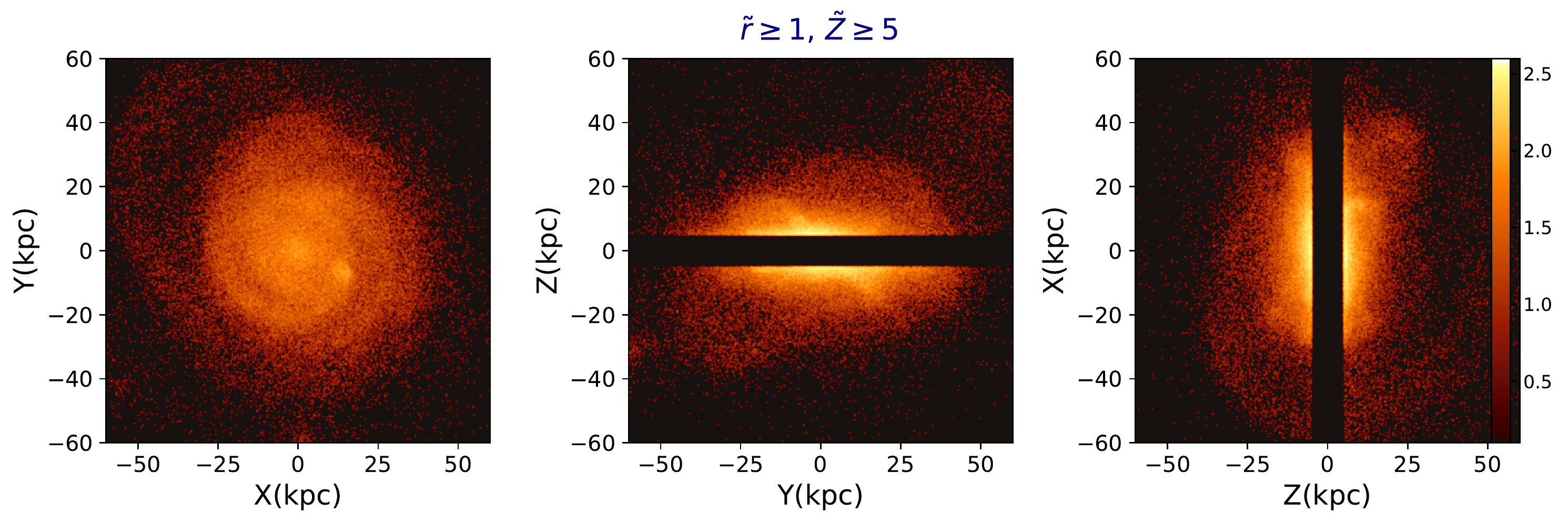}
\includegraphics[width=0.995\textwidth]{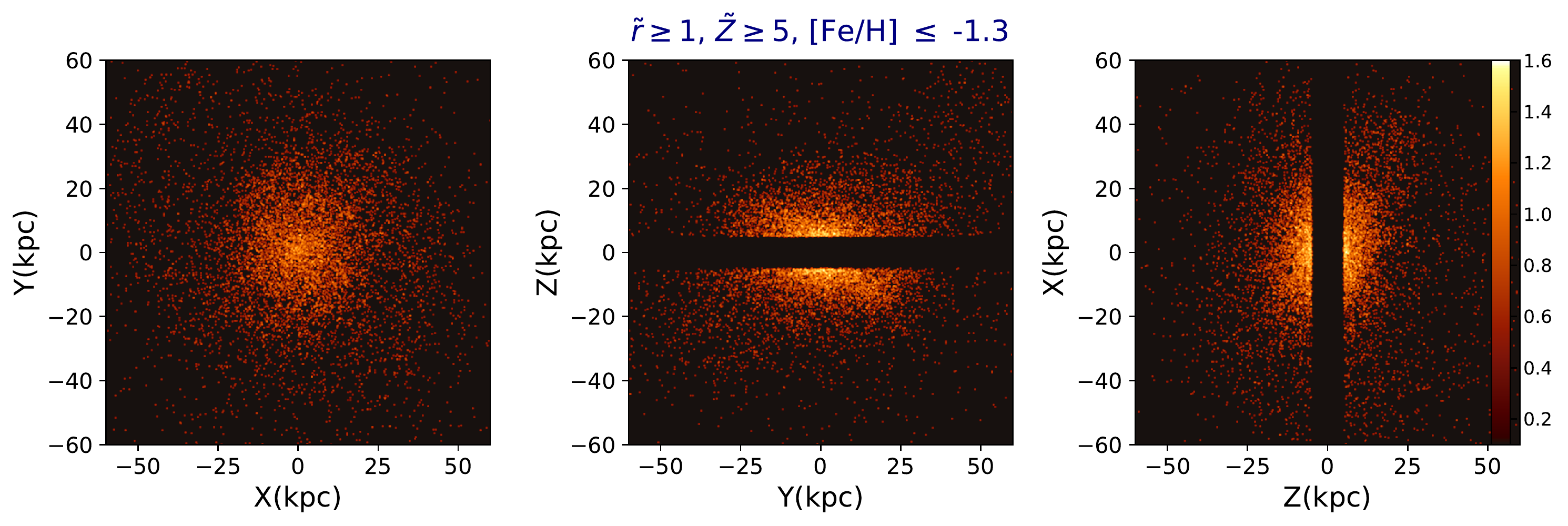}
\caption{Face-on (left) and (two) edge-on (middle-right) projections of stars in galaxy 1 in our sample. From top to bottom rows in the image present the spatial distribution of stars from $\tilde{r} \geq 1$, $\tilde{r} \geq 1 , \epsilon \leq 0.7$, $\tilde{r} \geq 1, \tilde{Z} \geq 5$ and  $\tilde{r} \geq 1, \tilde{Z} \geq 5,$ [Fe/H] $\leq$ -1.3, respectively. The color bars refer to the surface number density of stars in different projections.}
\label{disk-decomposed-star}
\end{figure*}

Before we analyse the $\beta$ radial profile above in depth, here we present a spatial map of stars inferred from different selection cuts. This enables us to develop an intuition on the impact of different selection criteria on the spatial distribution of stars. Figure \ref{disk-decomposed-star} presents an image for a typical galaxy, galaxy 1, along with three different projections. Top to bottom rows present the 2D surface density map of stars with $\tilde{r} \geq 1$, $\tilde{r} \geq 1 , \epsilon \leq 0.7$, $\tilde{r} \geq 1, \tilde{Z} \geq 5$ and  $\tilde{r} \geq 1, \tilde{Z} \geq 5,$ [Fe/H] $\leq$ -1.3, respectively. Here $\tilde{X} \equiv X/$kpc with $X = (r,Z)$. Interestingly, the kinematic cut of $\epsilon \leq 0.7$ and the spatial cut of $\tilde{Z} \geq 5$ both decrease the spiral structure of the galaxy in the disk plane. While, the current metallicity cut diminishes the population of stars substantially, as [Fe/H] $\leq -1.3$ sits on the tail of the stellar distribution. 

Having fully described the details of different  lines as presented in Figure \ref{beta-radial-profile}, below we analyze the $\beta$ radial profile inferred from each of them in depth and make a halo classification based on the behavior of $\beta$ radial profile.

\subsubsection{$\beta$ inferred halo classification from selection criteria}
\label{Halo-cat}
Having presented the $\beta$ radial profile for halos in our sample, here we make a halo classification based on the sensitivity of $\beta$ radial profiles to different selection criteria. The goal is to quantify the sensitivity of $\beta$ to different stellar orbital types achieved from different selections. 

In our analysis, for every galaxy, we use the following empirical classifier:
\begin{equation}
\label{classifier}
\delta \beta \equiv  \frac{ \sum_{i} | \beta_i^{\mathrm{Fe}} - \beta_i^{\mathrm{No-cut}}| } { \sum_{i} \mid \beta_i^{\mathrm{No-cut}}\mid}.
\end{equation}
where $\beta_i \equiv \beta(r_i)$ is used for the sake of brevity. In addition, in Eq. (\ref{classifier}), $\beta^{\mathrm{Fe}}_i$ refers to the hybrid selection of $\mathrm{[Fe/H]}<-1.3$ with $|Z| \geq 5$ \rm{kpc}, while $\beta_i^{\mathrm{No-cut}}$ stands for the case with no spatial cut, as described above. The selection of $\beta^{\mathrm{Fe}}_i$ and $\beta_i^{\mathrm{No-cut}}$ in the halo classifier is owing to the visual differences seen in the curves associated with them, as presented in Figure \ref{beta-radial-profile}. 
\floatstyle{boxed} 
\begin{figure*}[!]
\center
\includegraphics[width=0.99\textwidth]{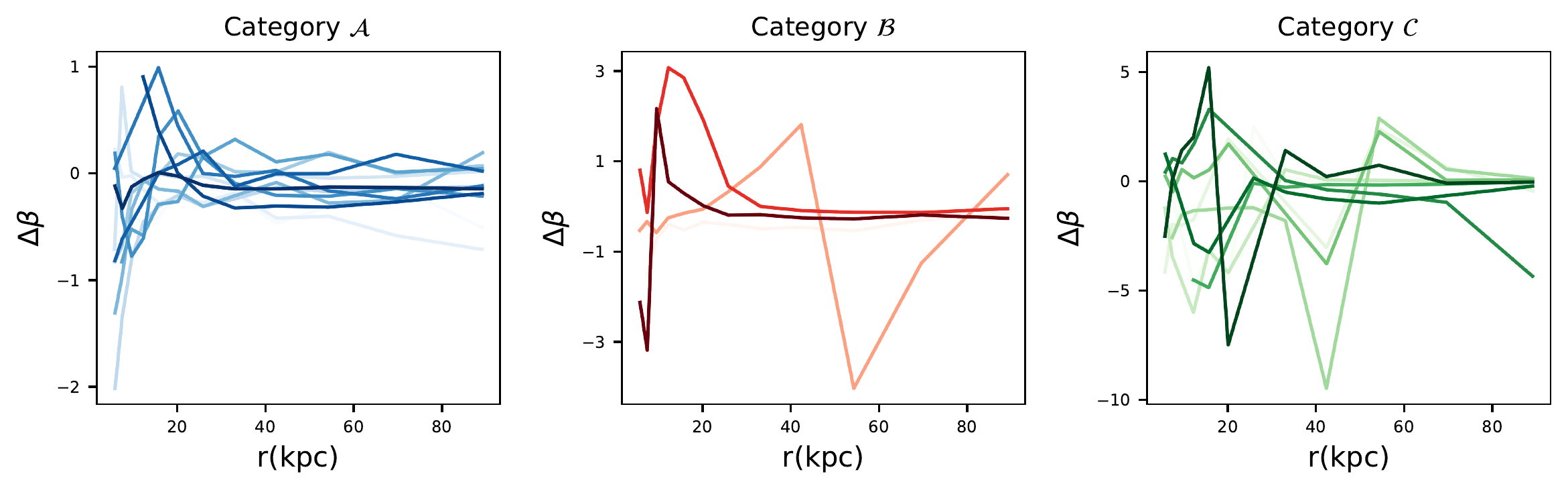}
\caption{The radial profile of $\Delta \beta(r) \equiv 
\left(\beta^{\mathrm{Fe}}/\beta^{\mathrm{No-cut}}\right)(r) -1 $ for halos from different categories. As stated in Section \ref{Halo-cat}, halos from Categories $\mathcal{A}$ through $\mathcal{C} $ show little to very strong response on different selection criteria for $\beta$, respectively. In addition, it is inferred that on average, halos from category $\mathcal{A}$ show smoother radial profile than the other two classes. This is more clarified in Figure \ref{radial-beta-detail}.}
 \label{classes}
\end{figure*}
As a next step, we define distinct halo classes based on the magnitude of $\delta \beta$ as listed below: 

$\bullet$ \textit{Category} $\mathcal{A}$:
$\delta \beta \leq 0.35$. Halos part of this category show the least sensitivity to different selection criteria. There are 12 halos in this category, including halos [1,4,5,6,7,13,15,16,17,18,22,23]. This shows $\beta$ is a very robust quantity in about 48\% of halos in our sample.

$\bullet$ \textit{Category} $\mathcal{B}$: 
$ 0.35 \leq \delta \beta \leq 0.70$. Halos in this class, show a mild response to different selection criteria. There are 4 halos in this class, including halos [10, 12, 21, 25]. 

$\bullet$ \textit{Category} $\mathcal{C}$: $ \delta \beta \geq 0.70$. Halos belong to this class show a strong dependency to different selections in $\beta$ calculation. There are in total 9 halos in this class, including halos [2,3,8,9,11,14,19,20,24]. This demonstrates that 
in 36\% of halos in our sample, $\beta$ is not a robust quantity as it depends on different selection criteria. 

Figure \ref{classes} presents the radial profile of the fractional difference between $\beta^{\mathrm{Fe}}_i$ and $\beta_i^{\mathrm{No-cut}}$ for halos from different classes. In each case, we present:
\begin{equation}
\label{frac-change}
\Delta \beta(r) \equiv 
\left( \frac{\beta^{\mathrm{Fe}}(r)}{\beta^{\mathrm{No-cut}}(r)}\right) -1.
\end{equation}
eliminating the points for which the denominator gets below 0.05. From the plot, it is inferred that halos belong to category $\mathcal{A}$ in average show smoother radial profile than those from $\mathcal{C}$. Below, we make this point more clear. 

\subsubsection{Detailed radial profile of $\beta$}
\label{beta-behavior}
Having identified various halo classes, in what follows, we study the $\beta$ radial profile in some depth and make a connection to the aforementioned halo categories. Considering different profiles, we are dealing with the following behaviors: 
\begin{figure*}[!]
\center
\includegraphics[width=0.98\textwidth]{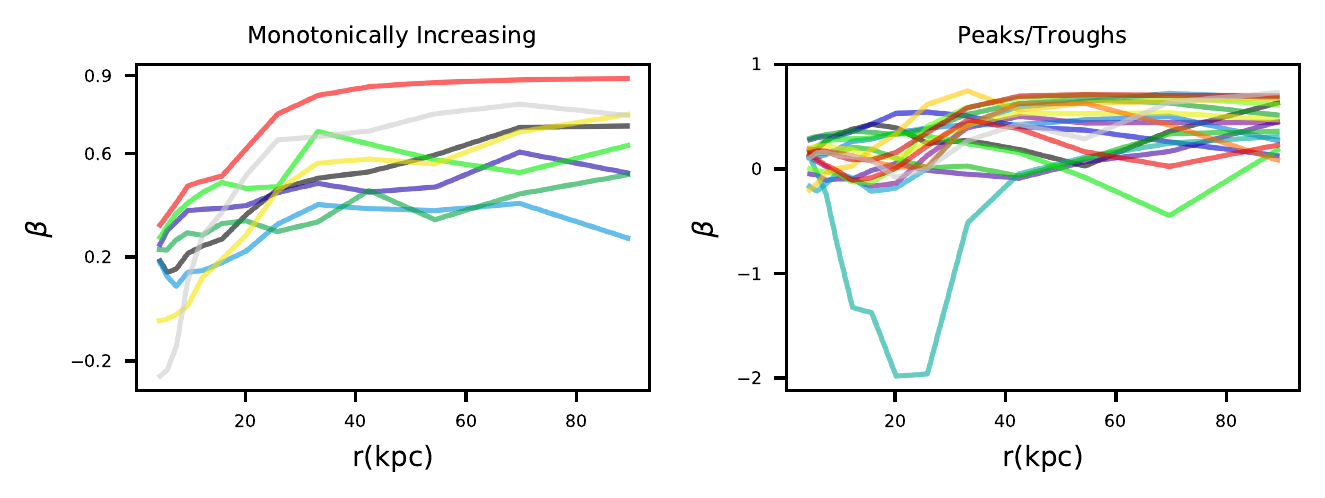}
\caption{ (Left panel) the profile of $\beta$ with monotonically increasing radial profile. 32\% of halos are in this category. Almost all of them are from category $\mathcal{A}$, with one exception belong to class 
 $\mathcal{B}$. (Right panel) the $\beta$ radial profile for cases with peaks and troughs. About 68\% of halos are belonged to this category from which only 29\% of halos are belong to category $\mathcal{A}$. This means that care must be taken in interpreting the observed peaks/troughs in the $\beta$ radial profile as it might be an indication of different selection criteria. }
\label{radial-beta-detail}
\end{figure*}
$\bullet$ \textit{Monotonically increasing:} From Figure \ref{beta-radial-profile}, it is inferred that in 32\% of halos in our sample, $\beta$ increases with the radius implying that farther out from the center, the orbits become progressively more radial (i.e., less tangential) and thus less rotationally-dominated. This is in agreement with e.g. \citet[][]{2017ApJ...835..193E}. In such cases, $\beta(r)$ profiles are monotonically increasing (i.e., galaxies 6, 7, 13, 15, 18, 22, 23 and 25). There are no visible peaks or troughs in the $\beta$ radial profile. Such galaxies are also not sensitive to our various selection criteria (the various lines overlap). So $\beta$ radial profile is less biased at identifying distinct stellar orbital types. Comparing the halos from this class with the above halo categories, it is evident that almost all of the halos in this case are from category $\mathcal{A}$. The only exception to this is halo 25 which is part of class $\mathcal{B}$. 

$\bullet$ \textit{Profile with Peaks/Troughs:} On the contrary, in 68\% of cases (e.g., galaxies 1, 2, 3, 4, 5, 8, 9, 10, 11, 12, 14, 16, 17, 19, 20, 21 and 24), $\beta(r)$ experience visible peaks and troughs, in line with the recent observations \cite[e.g.][]{2012ApJ...761...98K,2019ApJ...879..120C}. 
About 29\% of halos in this class are belong to category $\mathcal{A}$ while the rest are from classes $\mathcal{B}$ and $\mathcal{C}$. As the vast majority of halos with peaks/troughs show mild/strong response to different selection criteria, care is warranted because the presence or absence of such peaks and troughs might actually depend on different selection criteria. For instance, selecting stars with $\mathrm{[Fe/H]} < -1.3$ typically results in $\beta(r)$ profiles without these features. Figure \ref{radial-beta-detail} compares the details of $\beta$ profiles in each of the aforementioned cases. On the left panel, we show the $\beta$ radial profile for monotonically increasing cases, while on the right panel, we present the cases with peaks/troughs. 

\subsection{Impact of different metallicity cuts on $\beta$ radial profile} 
\label{beta-metal}
Having presented the $\beta$ radial profile with the LAMOST based metallicity cut, here we generalize such a selection criteria and study the impact of changing different metallicity cuts on the $\beta$ radial profile. As already stated in Sec. \ref{age-metal}, since the stellar metallicity is negatively correlated with the stellar age, various metallicity cuts somewhat choose stars with different ages. It is then very intriguing to study the impact of the stellar age on the velocity anisotropy. 

\begin{figure*}
\center
\includegraphics[width=0.99\textwidth]{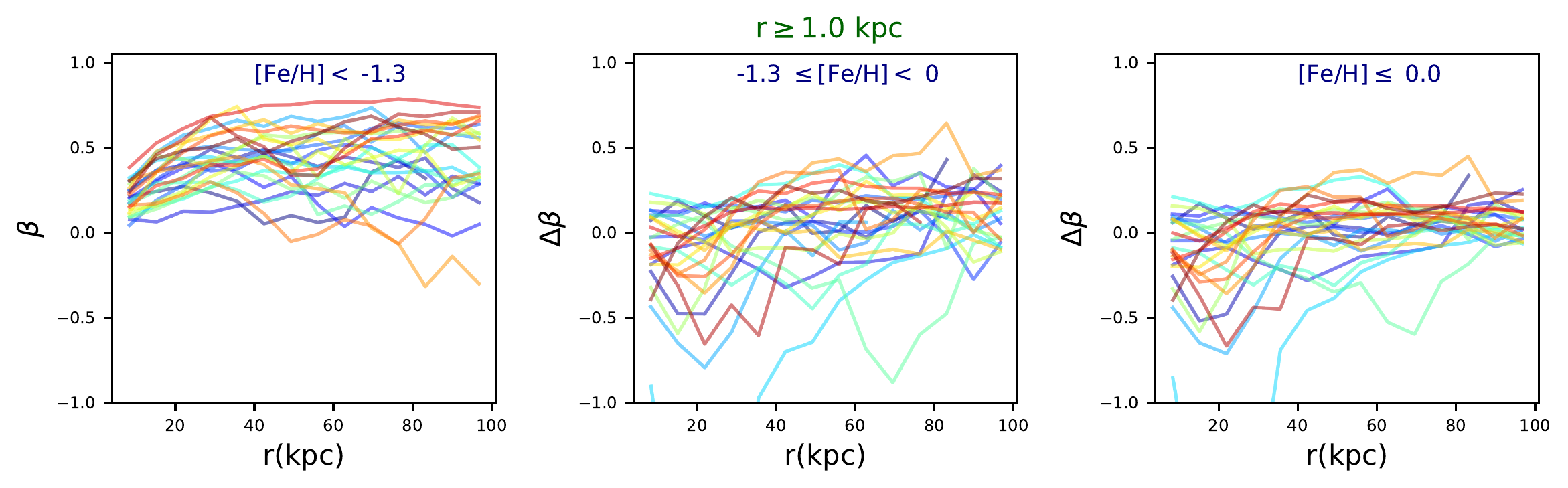}
\includegraphics[width=0.99\textwidth]{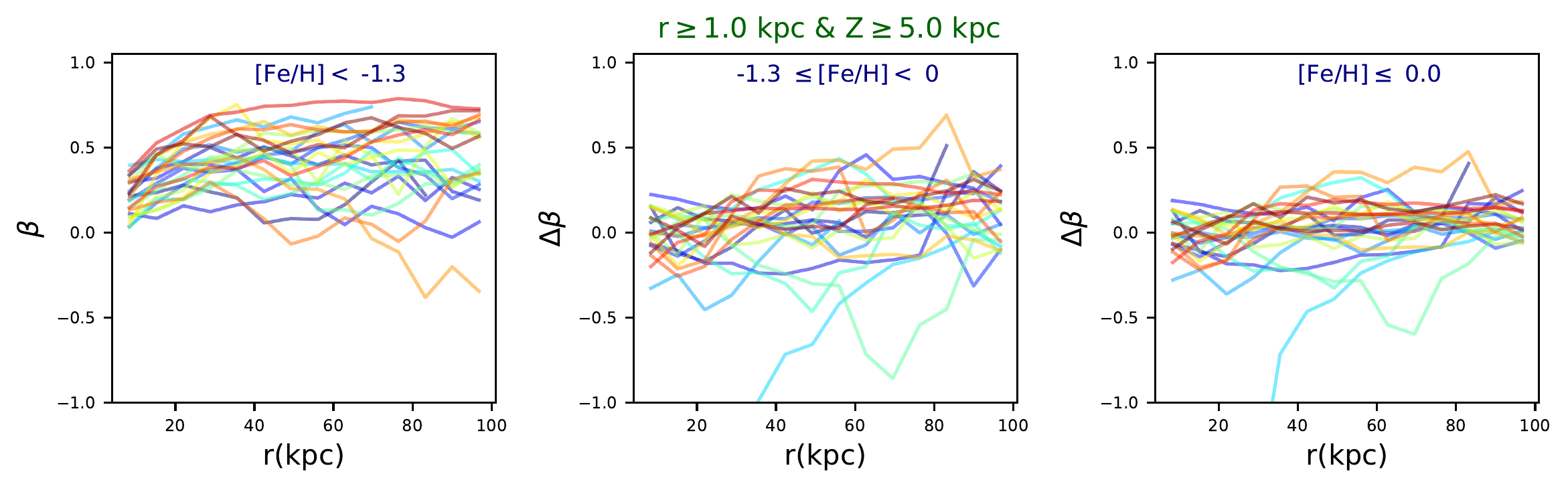}
\caption{ $\beta$ radial profile for stars located at $r \geq 1$ kpc (first row) as well as those with $r \geq 1$ kpc $\& Z \geq 5$ kpc (second row). In each row, from the left to the right, we consider stars with [Fe/H] $\leq -1.3$, -1.3 $\leq$ [Fe/H] $\leq$ 0 and [Fe/H] $\leq$ 0.0. While in the left panels we present the $\beta$ itself, to facilitate the comparison, in the middle and right panels, we compute the $\delta \beta$ subtracting the $\beta$ from the left panel with [Fe/H] $\leq -1.3$. It is clearly seen that expanding over the range of metallicity, on-average, diminishes the $\beta$. }
 \label{Age-Metal-beta}
\end{figure*}

Figure \ref{Age-Metal-beta} presents the $\beta$ radial profile for stars with different spatial cuts vs various metallicity criteria. The upper panel presents the $\beta$ profile for stars located at $r \geq 1$ kpc, while the bottom one shows the $\beta$  profile for stars located at $r \geq 1$ kpc and $Z \geq 5$ kpc. In each row, from the left to the right, we expand over the metallicity cut, from [Fe/H] $\leq -1.3$, -1.3 $\leq$ [Fe/H] $\leq$ 0 to [Fe/H] $\leq$ 0.0, respectively. In line with the conclusion of Sec. \ref{age-metal}, we might expect that chosen stars in the left panel are on average older than those in the middle one, while stars on the right panel are the mixture of the young and old stars. It is inferred that the left panels have higher $\beta$s than the middle ones. This makes sense, as old stars are less rotationally supported implying lower $\sigma_{\theta}$ and $\sigma_{\phi}$ and thus higher $\beta$. On the contrary, young stars have higher tangential velocity components which correspond to lower values of the $\beta$.  Finally,  mixed stars, on the right panel, have $\beta$s in between the young and old populations. Interestingly, such a conclusion is almost independent of the spatial cut, as the top and bottom rows are generally quite similar. 

\begin{figure*}[t!]
\center
\includegraphics[width=0.99\textwidth]{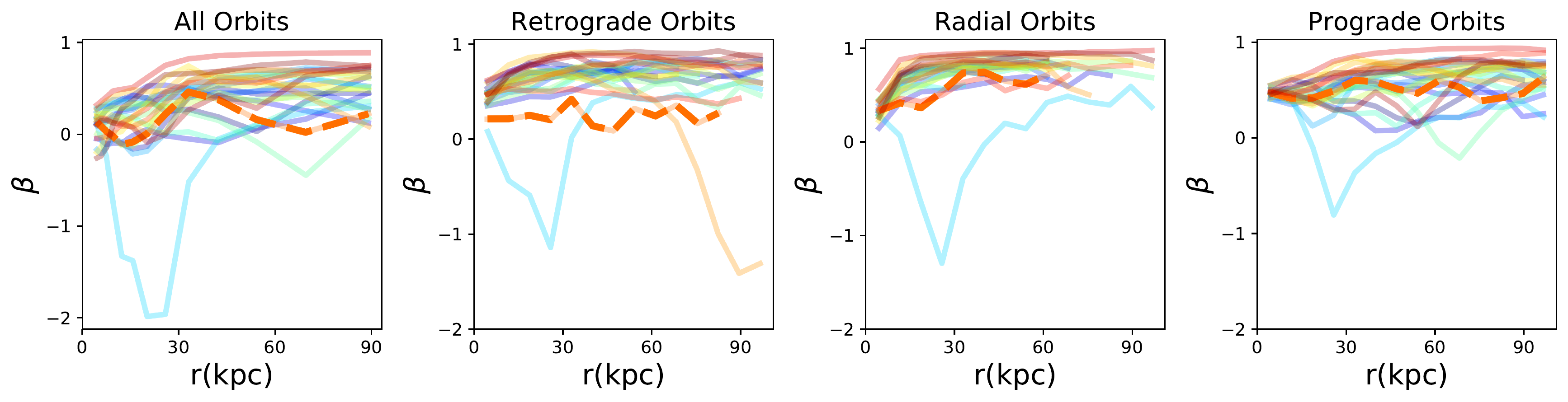}
\caption{$\beta$ profile for stellar on different orbital types. From the left to right, we study the $\beta$ radial profile for all, retrograde, radial and prograde orbits, respectively. It is generally seen that stars on radial orbits have slightly higher $\beta$ than the retrograde and the prograde orbits. 
On the contrary, stars on retrograde and prograde orbits show similar trends in $\beta$ radial profile. }
 \label{beta-orbit}
\end{figure*}

\subsection{Impact of different orbital types on the $\beta$ radial profile}  \label{beta-orb}
Next, we explore the impact of stellar orbital types on the $\beta$ radial profile. Figure~\ref{beta-orbit} shows $\beta(r)$ profiles for the retrograde, radial and prograde orbits (plus the entire stellar set on the left-hand panel). In each case, we compute $L^i_z \equiv \vec{L}^i \cdot \vec{\hat{L}}_{\mathrm{tot}}$ for $i$-th star, where $\vec{L}^i $ refers to the angular momentum of the $i$-th star while the $\vec{\hat{L}}_{\mathrm{tot}}$ describes the unit vector along with the total angular momentum of stars. We define our three stellar orbital types as follows \citep{2020ApJ...901...48N}: 

$a.$ Retrograde orbits: $L^i_z \leq -500~\mathrm {kpc \times km ~ sec^{-1}} $,

$b.$ Radial orbits:  $ -500 \leq L^i_z/\left( \mathrm {kpc \times km ~ sec^{-1}} \right) \leq 500$,

$c.$ Prograde orbits:  $ L^i_z \geq 500 ~\mathrm {kpc \times km ~ sec^{-1}}$.

While there are some variations on the exact behavior of individual galaxies, it is generally true that stars on radial orbits have slightly higher $\beta$ than the retrograde and the prograde orbits. That is expected as radial orbits have generally higher $\sigma_r$ and thus their $\beta$ would be higher based on Eq. (\ref{beta}). On the contrary, stars on retrograde and prograde orbits seem to have similar trends though stars on prograde orbits have slightly higher $\beta$s than the retrograde ones. 

\begin{figure*}
\center
\includegraphics[width=0.99\textwidth]{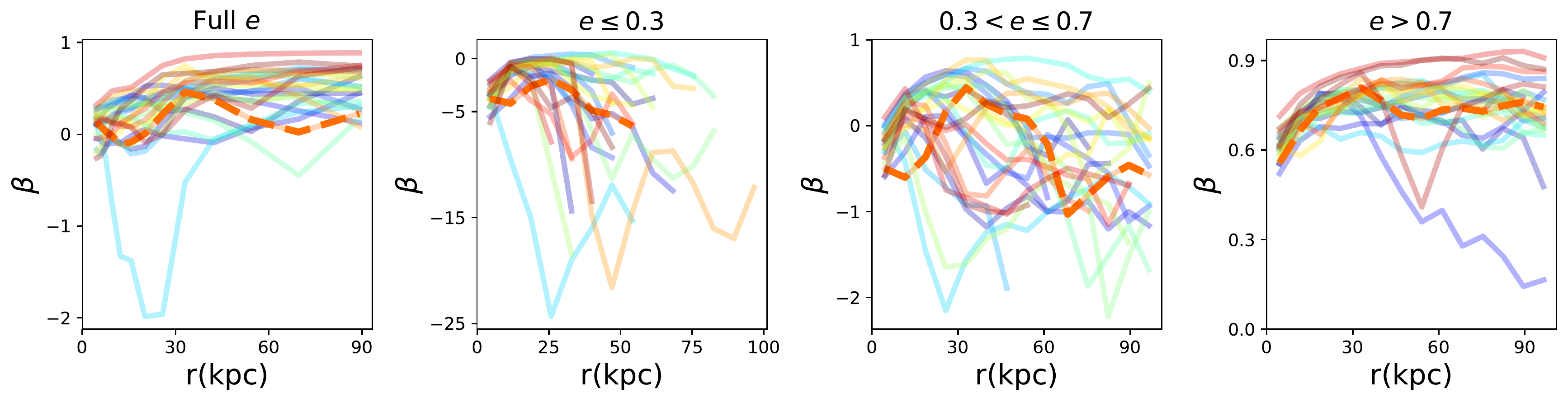}
\caption{$\beta$ profile for different eccentric orbits. From the left to right we present the entire of stars, stars with $e \leq 0.3$, $ 0.3 \leq e \leq 0.7 $ and $e \geq 0.7$, respectively. It is noted that 
low-eccentricity stars have lower $\beta$ than the parent population as such stars are on nearly-circular orbits, On the other contrary, stars with higher eccentricities are more on radial orbits than tangential, producing higher $\beta$ values.  
}
 \label{beta-eccentricity}
\end{figure*}

\begin{figure*}
\center
\includegraphics[width=0.99\textwidth]{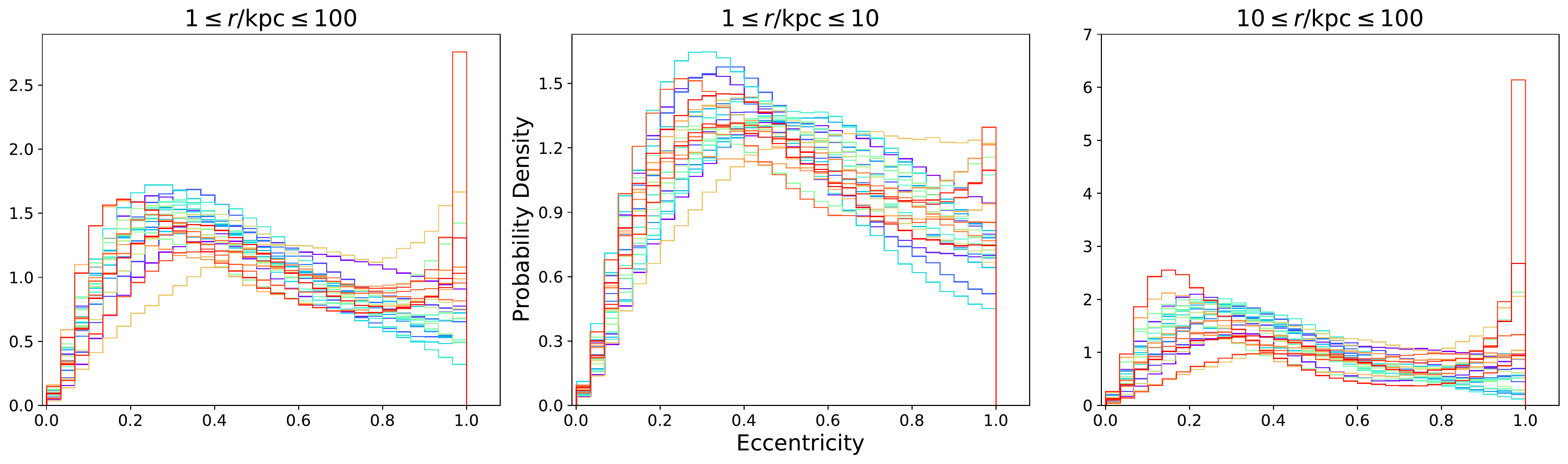}
\caption{1D distribution of the eccentricity of stars in our galaxy sample. Quite interestingly, most galaxies have an eccentricity peak between 0.2-0.4. In very few cases, there is a second peak at higher eccentricities of about 1.0 coming from those stars located at larger distances.}
\label{e-dist}
\end{figure*}

\subsection{Impact of eccentricity on the $\beta$ radial profile} \label{beta-ee}
As the next case, here we analyze the impact of the orbital eccentricity on the $\beta$ radial profile. Where the eccentricity is approximately defined as,

\begin{equation}
\label{eccentricity}
e \simeq \sqrt{1 - \left(\frac{L}{L_{\mathrm{circ}}}\right)^2},
\end{equation}
where $L_{\mathrm{circ}} \simeq \sqrt{G M(r_c) r_c}$ with $M(r_c)$ referring to the total mass interior to $r_c$ describing the circular radii, see Equation 3 in \cite{2015ApJ...814...57M} and  \cite{2021ApJ...913...36E} for more details. 

 We note that Eq. \ref{eccentricity} is an approximate expression for the orbital eccentricity. This together with the above estimation for $L_{\mathrm{circ}}$ gives rise to a negative result inside the square root in Eq. \ref{eccentricity} for a very small population of stars. As this is a consequence of a breakdown in the above assumption for that population of stars, we remove them from our sample. Finally, to check that Eq. ref{eccentricity} gives rise to a reasonable result for the eccentricity, we took a slightly different approach, \cite[see for instance Eqs. 1, 2 of][]{2019MNRAS.482.3426M}. While the second scheme also fails for stars with a pericenter less than the softening length of the TNG simulation in which we do not have a well-defined expression for the gravitational potential, our results in the overlapping region indicate a satisfactory correspondence. Owing to this, in what follows, we use Eq. \ref{eccentricity} and infer the eccentricity accordingly.

Figure \ref{beta-eccentricity} presents $\beta(r)$ profiles for the entire stellar population (left column) and three eccentricity-selected sub-samples: $e \leq 0.3$, $0.3 \leq e \leq 0.7$ and $e \geq 0.7$. We note that low-eccentricity stars have much lower $\beta$ than the parent population. This is because these stars have nearly-circular orbits, with $\sigma_r$ being much lower than the $\sigma_{\theta}$ and $\sigma_{\phi}$. On the other hand, stars with higher eccentricities are more on radial orbits than tangential, producing higher $\beta$ values.  

Since the above eccentricity cut led to a substantial variation in the $\beta$ radial profile, to explore its impact further, in Figure \ref{e-dist}, we draw the 1D probability distribution function of eccentricity; from the left to right, we analyse the distribution for the entirety of stars, for those restricted to radii $ 1 \leq r/\mathrm{kpc} \leq 10$ and for stars in the range $ 10 \leq r/\mathrm{kpc} \leq 100$, respectively. 
From the plot, it is evident that in most cases, the eccentricity peaks between 0.2-0.4. Furthermore, there is a signal of bi-modality in the eccentricity distribution. More explicitly, while in lower distances, the eccentricity is more concentrated on lower values, at large radii it shows some levels of bi-modality in which we have two peaks in the eccentricity distribution; one at relatively small radii and another one in very high values. 

\begin{figure*}
\center
\includegraphics[width=0.99\textwidth]{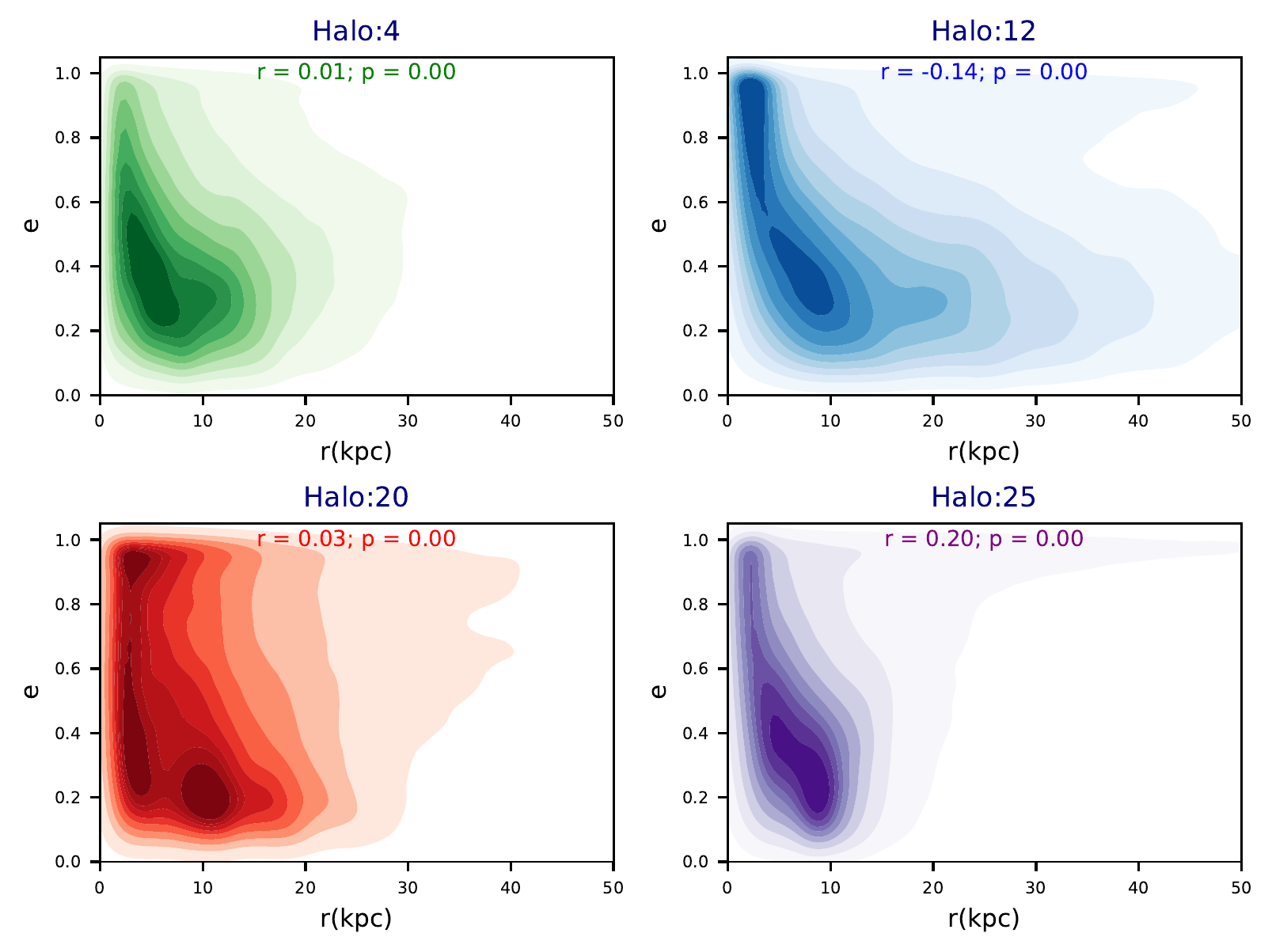}
\caption{2D distribution plot of the radii vs the eccentricity for a sub-sample four galaxies from our galaxy sample. The Spearman correlation demonstrates that the eccentricity is uncorrelated with the radii. 
}
\label{Eccentricity-R-dist}
\end{figure*}

Finally, Figure \ref{Eccentricity-R-dist} presents the 2D distribution for the radial dependency of the eccentricity for a sub-sample 4 out of our galaxy sample. In each plot, we also show the Spearman correlation between the radius and the eccentricity. It is clearly seen that these two quantities are un-correlated. Furthermore, different galaxies have their eccentricity peaks at different values and locations. For example, while galaxies 4 and 25 have a standard eccentricity peak of 0.2-0.5 in $r \leq 10$ kpc, galaxies 12 and 20 show some levels of bi-modality in the radial dependency of the eccentricity. It is seen that in these galaxies there is a very high eccentric distribution at very low radii, followed by a second peak at $r \leq 10$ kpc. Perhaps such a bi-modal distribution in the eccentricity is originated by galaxy mergers. However, this hypothesis must be validated and we leave this to future work.

\subsection{The $\beta$ energy profile} 
\label{beta-energy}
Since energy is an integral of motion, the $\beta$-energy profile, $\beta_\mathrm{E}$, may measure the velocity anisotropy in a less biased manner. It is then intriguing to compare the $\beta$ radial (in this section referred as $\beta_{\mathrm{r}}$) and energy profiles against each other and check their similarities.

\begin{figure*}
\center
\includegraphics[width=0.99\textwidth]{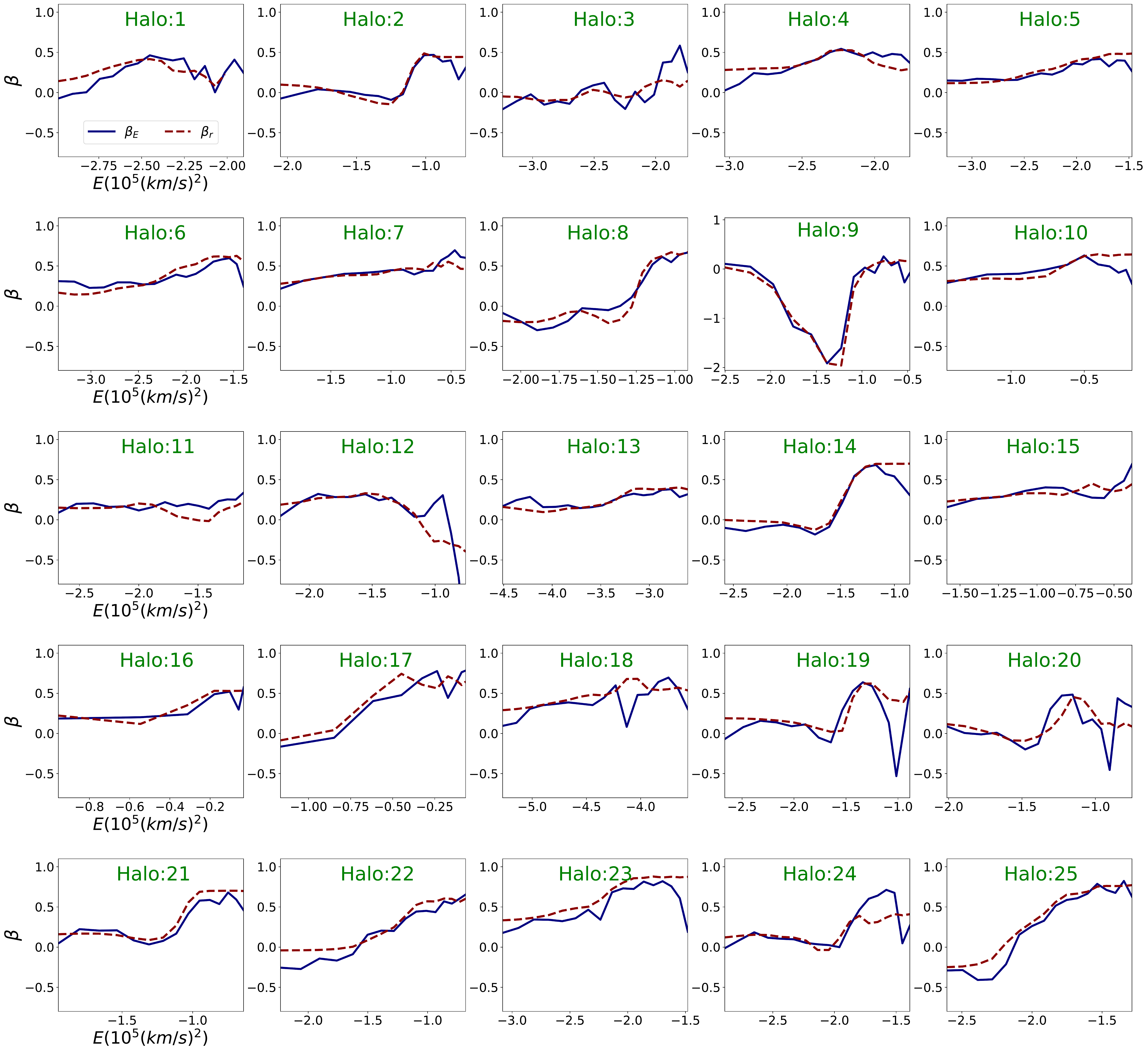}
\caption{$\beta$ energy profile vs the $\beta$ radial profile for different halos in our galaxy sample. For each energy shell, we infer the median distance of stars and compute the $\beta_\mathrm{r}$ for that location. In inferring the $\beta_\mathrm{E}$, we use a spatial cut of $1 \leq  r \leq 100 $ kpc.  }
 \label{beta-energy-profile}
\end{figure*}

Figure \ref{beta-energy-profile} shows the $\beta_\mathrm{E}$ profile for different galaxies in our sample. To infer $\beta_\mathrm{E}$, the stellar energy is read directly from the TNG outputs. We then generate some energy bins, from the energy min to its max value, and compute $\beta_\mathrm{E}$ for each of these individually. As the energy range differs from one galaxy to the other, the energy min and max are not the same for different galaxies. Furthermore, to ensure that in computing the $\beta_\mathrm{E}$, we use almost similar stars used in computing the $\beta_\mathrm{r}$, we imply a spatial cut of $ 1 \leq r \leq 100$ kpc. Overlaid on each panel, we also present the $\beta_\mathrm{r}$ profile inferred for the median distance in each energy shell. More explicitly, in every energy shell, used in computing the $\beta_\mathrm{E}$ profile, we read off the median of the stellar distance to the center and compute the $\beta_\mathrm{r}$ at that particular position. In each panel, the solid-blue line refers to the $\beta_\mathrm{E}$, while the dashed-red line describes the $\beta_\mathrm{r}$ profile. The overall agreement between the $\beta_\mathrm{E}$ and $\beta_\mathrm{r}$ is remarkable. There are however some differences between these profiles that are worth to be quantified. For this purpose, we write an empirical estimator
\begin{equation}
\label{classifier-ER}
\delta \beta_{\mathcal{ER}} \equiv  \frac{ \sum_{i} | \beta^i_{\mathrm{E}} - \beta^i_{\mathrm{r}}| } { \sum_{i} \mid \beta^i_{\mathrm{E}}\mid}.
\end{equation}
Next, in a way similar to what was done in Sec. \ref{Halo-cat}, we make distinct halo classes based on the magnitude of $\delta \beta_{\mathcal{ER}}$. We make 3 halo categories as listed below:  

$\bullet$ \textit{Category} $\mathcal{A_E}$:
$\delta \beta_{\mathcal{ER}} \leq 0.35$. The beta radial and energy profiles for halos part of this class are quite similar. There are 17 halos in this category, including halos [4,5,6,7,8,9,10,13,14,15,16,17,18,21,22,23,25]. Comparing the halos in this category with the ones part of $\mathcal{A}$, it is seen in 65\% cases, the halo from class $\mathcal{A}$ is also part of class $\mathcal{A_E}$.

$\bullet$ \textit{Category} $\mathcal{B_E}$: 
$ 0.35 \leq \delta \beta_{\mathcal{ER}} \leq 0.70$. Halos in this class, show a mild difference from the radial to energy profile. There are 6 halos in this class, including halos [1, 2, 11, 12, 19, 24]. From this list only one case is part of $\mathcal{B}$,  one is from  $\mathcal{A}$ while the rest are part of class $\mathcal{C}$. 

$\bullet$ \textit{Category} $\mathcal{C_E}$: $ \delta \beta_{\mathcal{ER}} \geq 0.70$. The $\beta_{\mathrm{E}}$ vs $\beta_{\mathrm{r}}$ in this class are not matched. There are 2 different halos part of this category, including [3, 20] and both of these halos are part of $\mathcal{C}$.

\section{Connection to observations}
\label{observations}
Having presented the $\beta$ radial profile using different cuts, here we make an in-depth comparison between the above theoretical outcomes and the actual observational results. There are two different aspects that could be addressed in our comparison. Firstly, the overall amplitude of the velocity anisotropy $\beta$ and secondly its radial shape. While the first case is easier to study, the second aspect is more challenging as it may depend on different selections/cuts employed for stars which is then less natural. Owing to this, we make the comparison for different selections, including $\tilde{r} \geq 1$, $\tilde{r} \geq 1 , \tilde{Z} \geq 5$ and $\tilde{r} \geq 1, \tilde{Z} \geq 5$, [Fe/H] $\leq$ -1.3. 

Below, we first point out few different observational studies. We then make a comparison between our results and different observations. 

\subsection{Observational constraints on $\beta(r)$} \label{observation-beta}
Observations measuring $\beta$ employ various approaches using different tracers, including K-giants, the blue horizontal-branch (BHB), RR Lyrae, and subdwarf stars in the solar neighbourhood. Below, we review few different observational studies.

\begin{figure*}
\center
\includegraphics[width=0.99\textwidth,trim = 6mm 2mm 0mm 1mm]{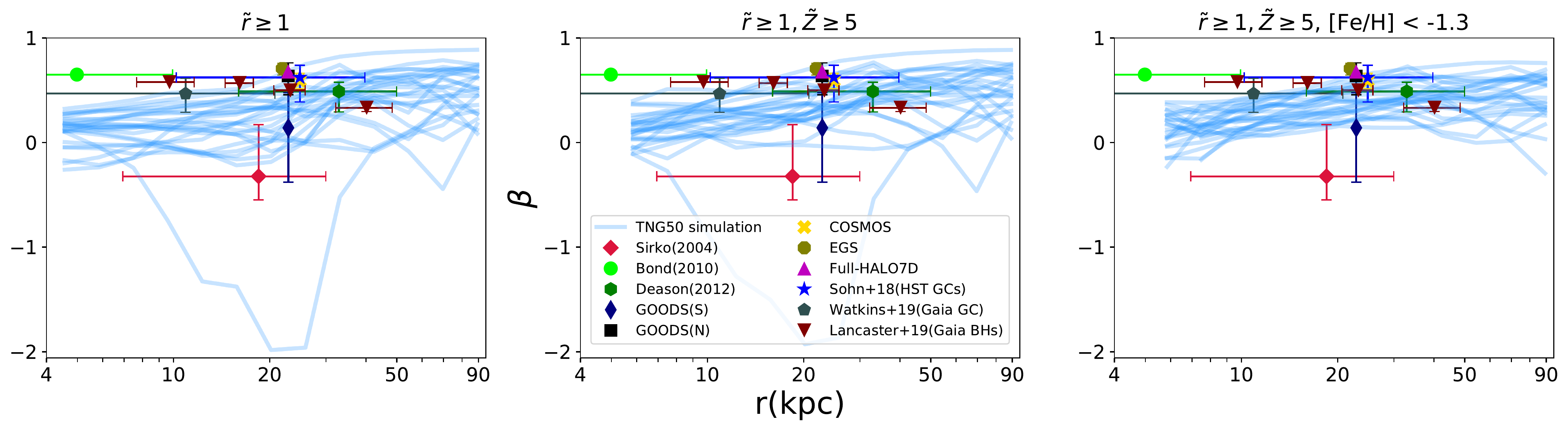}
\caption{Comparison between the $\beta$ radial profile inferred from the TNG50 simulation, using different selection criteria, and various isolated data points from different observations.
From the left to right, we present the $\beta$ profile with $\tilde{r} \geq 1$, $\tilde{r} \geq 1, \tilde{Z} \geq 5$ and 
$\tilde{r} \geq 1, \tilde{Z} \geq 5,$ [Fe/H] $\leq -1.3$, respectively. In each panel, we also present the data-points from \cite{2004AJ....127..914S, 2010ApJ...716....1B, 2012MNRAS.424L..44D, 2018ApJ...862...52S, 2019ApJ...873..118W, 2019MNRAS.486..378L, 2019ApJ...879..120C}. It is seen that adding the metallicity cut  with [Fe/H] $\leq$ -1.3, somewhat reduces the diversity of models as it removes the young stars, substantially. On the contrary, the model diversity is maximal for $\tilde{r} \geq 1$ in which the diversity is comparable to the error bars from changing the observations. This demonstrates that including/altering the stellar types may significantly affect the $\beta$, causing different observations to have different results!  
}
 \label{beta-obs-general}
\end{figure*}

\cite{2004AJ....127..914S} used a sample of blue horizontal-branch (BHB) stars from the SDSS survey and inferred the radial and tangential velocity dispersions in the radial distance $ 5 \leq r/\mathrm{kpc} \leq 30$. In Figure \ref{beta-obs-general}, we have converted these values to some estimators for $\beta$ in the aforementioned interval. 

\cite{2010ApJ...716....1B} used a large sample (18.8 million) of main-sequence stars derived from SDSS and POSS astrometry and estimated the $\beta \simeq 0.65$ in a radial range $ 3 \leq r/\mathrm{kpc} \leq 13$. 

\cite{2012MNRAS.424L..44D} 
used distant BHB stars with the Galactocentric distances 
in radial range
$16 < r/\mathrm{kpc} < 48$ as the kinematic tracers of the Milky Way dark halo and infer
$\beta \approx 0.5$, in line with local solar neighborhood studies.

\cite{2018ApJ...862...52S} used their HST proper motion (PM) measurements from 16 globular clusters (GCs) and find $\beta = 0.609 ^{0.13}_{-0.229}$ in the galactic distance range from $10-40$ kpc. 

\cite{2019ApJ...873..118W} used a sample of 34 halo GCs from Gaia data spanning in the radial range 2-21.1 kpc and estimated the velocity anisotropy $\beta = 0.46. ^{+0.15}_{-0.19} $.  

\cite{2019MNRAS.486..378L} assembled a very high purity set of 3064 BHBs using the spectroscopic data from SDSS as well as the astrometric data courtesy from the Gaia satellite. They computed the $\beta$, see the orange line in their Figure 9, in the galactic center radii range 7-50 kpc. Their extended data points allow us to probe the radial evolution of $\beta$. However, since it only covers 4 data points we present the results in our generic comparison in Figure \ref{beta-obs-general}.  

\cite{2019ApJ...879..120C} used the data from HALO7D (Halo Assembly in Lambda Cold Dark Matter: observation in 7 Dimensions) consisting of the KeckII/DEIMOS spectroscopic and the HST data applied for the main-sequence turnoff stars and estimate the $\beta$ for the full HALO7D sample as well as its individual fields, including  COSMOS, GOODS-N, GOODS-S and EGS. The details of the data-points and the $\beta$ values are given in Table 3 of \cite{2019ApJ...879..120C}. In Figure \ref{beta-obs-general} we present all of the results for the full-sample as well as individual ones. 

\begin{figure*}
\center
\includegraphics[width=0.99\textwidth,trim = 6mm 2mm 0mm 1mm]{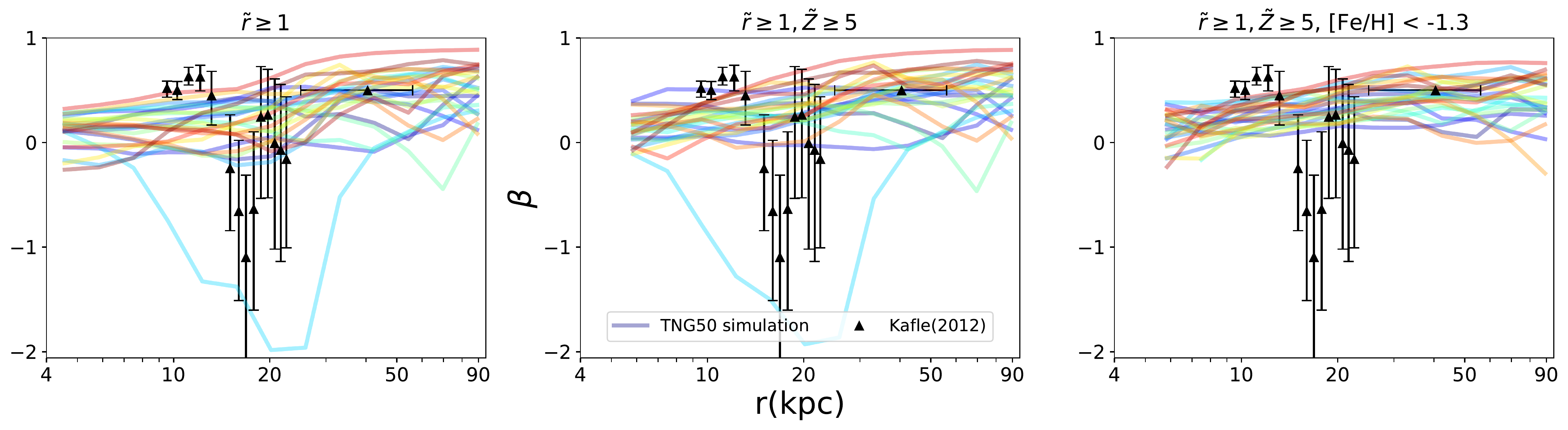}
\includegraphics[width=0.99\textwidth,trim = 6mm 2mm 0mm 1mm]{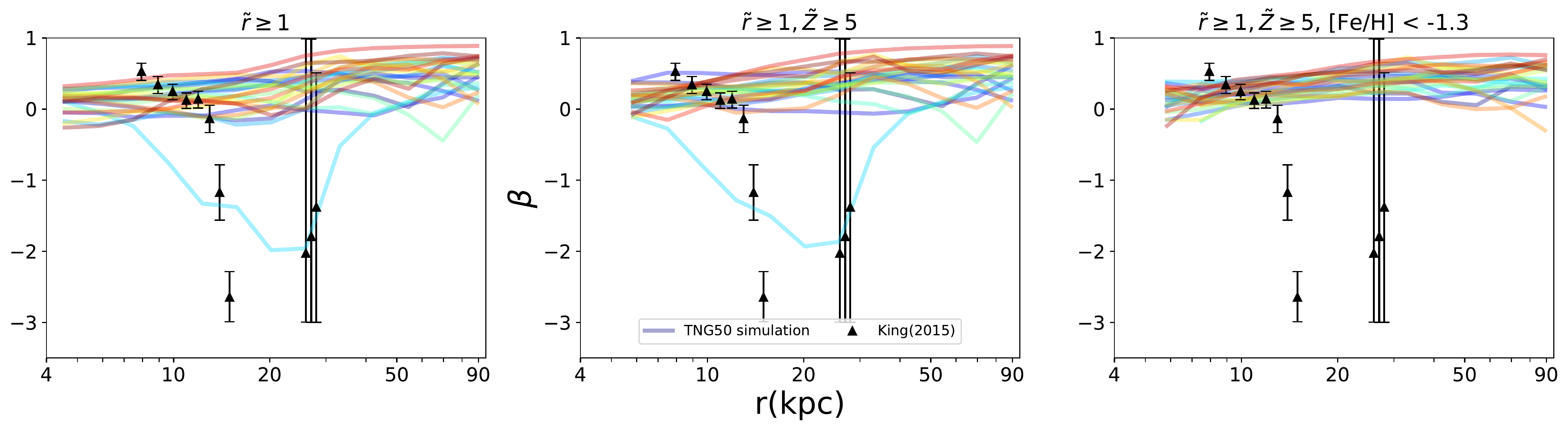}
\includegraphics[width=0.99\textwidth,trim = 6mm 2mm 0mm 1mm]{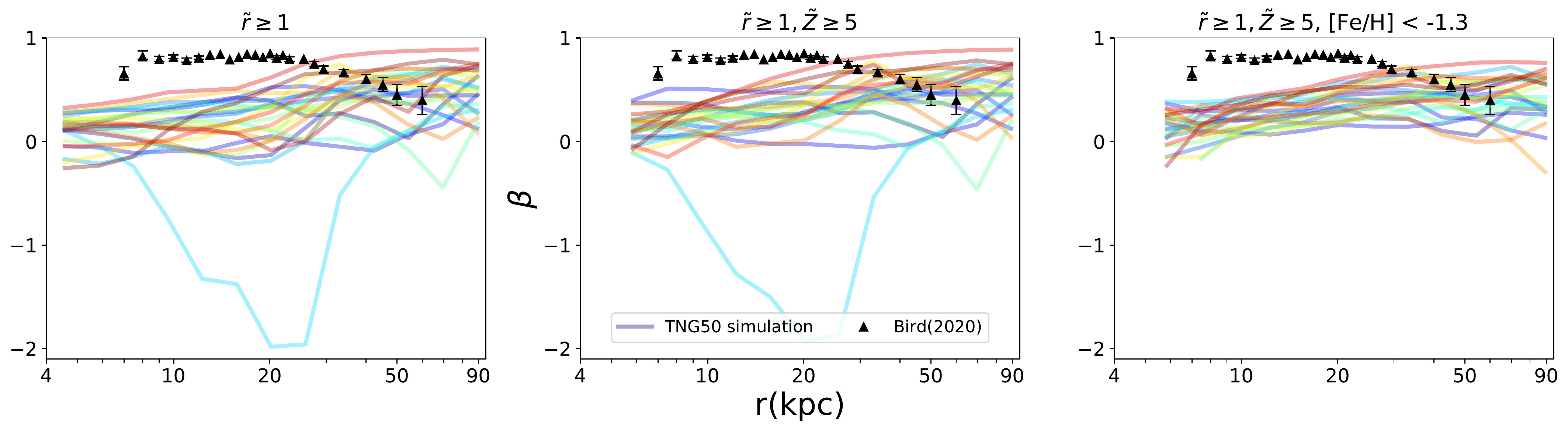}
\caption{Comparison between the $\beta$ radial profile from the TNG50 simulation with three extended observational data points. Top panel presents the data from \cite{2012ApJ...761...98K}, the middle panel shows the results from \cite{2015ApJ...813...89K} and the bottom one presents the results from \cite{2019AJ....157..104B}. In each row, we present the TNG50 results using different selection criteria. It is seen that there are a good overall agreement between the TNG50 results and these extended observations. The model diversity is somewhat comparable to the error bars in \cite{2012ApJ...761...98K}, while it is smaller than the ones from \cite{2015ApJ...813...89K}.  Finally, while the first two panels show a prominent dip in the $\beta$ radial profile, the third one does not show any particular dips. Since proper motion has not been considered in the first two cases, the third one may be more reliable.}
 \label{beta-obs-specific}
\end{figure*}

\cite{2012ApJ...761...98K} used a sample of 4664 BHB stars chosen from SDSS/SEGUE survey and measured the $\beta$ out to 25 kpc while estimating this up to 60 kpc. There are various trends seen in the data, including a smooth value of $\beta \sim 0.5$ in the inner part of the halo ($ 9 \leq r/\mathrm{kpc} \leq 12$), followed by a rather sharp falls off of $\beta$ in the range $r \sim 13-18$ kpc and a minimum of $\beta = -1.2$ at $r = 17$ kpc and an extra rising of $\beta$ at larger radii. At the outer parts of the halo, in the interval $25 \leq r/\mathrm{kpc} \leq 56$, the actual measurement of $\beta$ is not possible owing to the lack of the proper motions. They predicted the $\beta \sim 0.5$ in this interval. In figure \ref{beta-obs-specific}, top panel, we focus on these data and compare them with the TNG50 results obtained using different cuts. 

\cite{2015ApJ...813...89K} used a sample of 6174 faint F-type stars from the Hectospec spectrograph on MMT telescope plus a sample of 3330 BHB stars from SDSS and estimated the $\beta$ within 6-30 kpc of the MW. The results are presented in Table 3 of \cite{2015ApJ...813...89K}. In the middle panel of Figure \ref{beta-obs-specific}, we present their data and compare them against the simulation results obtained using different cuts. 

\cite{2019AJ....157..104B} made a sample of 7664 metal-poor K giants using LAMOST catalog and measured the $\beta$ at the galactocentric radii between 5-100 kpc. 
As they matched their line-of-sight velocities to the proper motions from the Gaia data, their final results might be less bias compared with former literature, such as \cite{2012ApJ...761...98K, 2015ApJ...813...89K}. In the bottom panel of Figure \ref{beta-obs-specific}, we present their results, comparing them against the TNG50 results obtained using different cuts.

\subsection{Comparison between the theory and observation} \label{comp-beta}
Here we make a comparison between the aforementioned observational results listed in Sec. \ref{observation-beta} and the outcome of the TNG50 simulation obtained from employing different selection criteria. 

Before we dig into the details of the comparison, we make some remarks regarding to the potential impact of substructures on the $\beta$ radial profile. Observationally, \cite{2019AJ....157..104B} showed that including the effect of the Sagittarius stream may lead to a dip in the $\beta$ radial profile. Consequently, the inclusion and the removal of the substructures are in principle very important. Theoretically, there have been some studies \cite[see for e.g][and references therein]{2018ApJ...853..196L} showing that the substructures may alter the shape of the $\beta$ radial profile. In the context of the TNG50 simulations, specifically, the substructures are all removed from the central halos and are only part of the group catalogue. As the entire focus on this paper is on the central halos, our analysis is safe of the direct impact of the substructures. Our analysis however is missing the indirect impact of substructures through their gravitational potential. This is much harder to quantify as we need to model their orbits which require greater time cadence in the outputs. We therefore ignore this effect in the current work and leave its analysis to a future study.

We split the comparison to two different sets; presenting them in two separate plots. 

In Figure \ref{beta-obs-general}, we present the individual data points from 7 different observational data sets \citep{2004AJ....127..914S, 2010ApJ...716....1B, 2012MNRAS.424L..44D, 2018ApJ...862...52S, 2019ApJ...873..118W, 2019MNRAS.486..378L, 2019ApJ...879..120C}. Adding the TNG50 results obtained from 3 different selection criteria, $\tilde{r} \geq 1$ (left panel), $\tilde{r} \geq 1 , \tilde{Z} \geq 5$ (middle panel) and $\tilde{r} \geq 1, \tilde{Z} \geq 5$, [Fe/H] $\leq$ -1.3 (right panel). From the plot, it is inferred that the model diversity is maximal for the left panel, while it is minimal in the right panel with metallicity cuts. This means that it is easier to find a theoretical model that matches more observations from the left panel than the right. 
Furthermore, despite the fact that there is not any single galaxy that matches all of different observational points, there are quite few of them that gets close to quite few observations, especially for the left and middle panels with no metallicity cuts. 
To summarize, the diversity of different realizations from the TNG50 are comparable with the differences between various observations. This means that TNG50 models can explain different observations.

In Figure \ref{beta-obs-specific}, we focus on three extended observations from \cite{2012ApJ...761...98K, 2015ApJ...813...89K, 2019AJ....157..104B} in which we put the data from three individual observations in different rows and make the comparison against the TNG50 results. Again, different panels refer to various selection criteria. The advantage of using the same observation is that we are generally less biased by different selection criteria from various observations. However, we are still affected by some systematic errors owing to the fact that the observational data of \cite{2012ApJ...761...98K, 2015ApJ...813...89K} were collected before Gaia. This is likely to introduce some biases owing to the lack of the actual treatment of the proper motion \cite[see e.g.][for more details]{2017ApJ...841...91H}. Consequently, the interpretation of the $\beta$ radial profile might not be straightforward. AS it is seen from the bottom row, the analysis of \cite{2019AJ....157..104B} did not lead to a strong dip in the $\beta$ radial profile.

From the plot, it is seen that there are few models that get very close to some data points, and within their error bars. Again, the model capabilities in reproducing the data slightly depend on different selection criteria. 

In summary, the TNG50 results are in overall good agreement with different observational points, though some of these are closer to the data. 

Since the diversity of TNG50 models diminish by considering a metallicity cut, we expect to have more population of models near the peak of agreement while less of them away from the peak. On the contrary, for models with a higher diversity, the chance to find a model realization that passes close to the rare data points are higher, though the population of models near the peak of agreement is less.  
\section{Conclusions}
\label{concl}

In this manuscript, we analyzed the stellar kinematics as traced with the velocity anisotropy, $\beta$, profile for a sample of 25 Milky Way like galaxies from TNG50 simulation. We made an in-depth study of the $\beta$ radial and the energy profiles, at redshift of $z=0$, exploring the impact of different selection criteria on the aforementioned quantities. 

Below, we summarize the main take-aways from this study: 

$\bullet$ It is observed that the stellar metallicity is negatively correlated with the stellar age, with a Spearman coefficient  $|r| \simeq 0.5-0.8$. Furthermore, it is also negatively correlated with the location of stars, meaning that older stellar population are mostly located in the galaxy outskirt while the younger ones are closer to the center. This is expected as the MW is still in its star-forming stage. Based on this, we explored the impact of different metallicity cuts on the velocity anisotropy of stars. 

$\bullet$ We inferred the $\beta$ radial profile using different selection criteria and made halo classification, categories $\mathcal{A}$-$\mathcal{C}$, based on the sensitivity of their profiles on various cuts. 

$\bullet$ Halos part of $\mathcal{A}$ class, show a very mild sensitivity to different selections, where $\delta \beta \leq 0.35$. And members of category $\mathcal{B}$ exhibit some levels of sensitivity, with $ 0.35 \leq \delta \beta \leq 0.7$, to different cuts. Finally, halos associated with $\mathcal{C}$ show a very strong dependencies, $\delta \beta \geq 0.7$, to model selections. We showed that there are 48\%, 16\% and 36\% of halos in each of these classes, respectively. 

$\bullet$ We analyzed the $\beta$ radial profile and demonstrated that it is divided to either monotonically increasing profiles (32\% of halos) or the ones with peaks/troughs (68\% of population). It is shown that, the $\beta$ profile from the monotonically increasing class smoothly increases from the interior to the exterior part of the halo. While, $\beta$ associated with the class with peaks/troughs experiences some local fluctuations. We further demonstrate that almost all of the halos from the first class are part of category $\mathcal{A}$, with only one exception being part of class $\mathcal{B}$. On the contrary, members of the class with peaks/troughs are mostly part of category $\mathcal{B}$ or $\mathcal{C}$, with only 29\% of them being part of category $\mathcal{A}$. This means that care should be taken in 
interpreting the observed peaks/troughs as they might depend on the actual selection criteria!

$\bullet$ We probed the impact of different metallicity cuts on the $\beta$ profile and demonstrated that older stars are less rotationally supported and thus have higher 
$\beta$. On the other hand, younger stars are more rotationally supported and have lower $\beta$. 

$\bullet$ We explored the impact of different  eccentricity based radial cuts on the $\beta$ radial profile. More explicitly, we divided stars to low, medium and high eccentric chunks. This explicitly showed that low eccentricity orbits correspond to lower $\beta$ than the high eccentricity orbits. This is expected as low eccentric orbits correspond to less radial velocity variance which naturally leads to smaller $\beta$.  

$\bullet$ We computed $\beta$ for different stellar types; demonstrating that stars on radial orbits have higher velocity anisotropy than those on prograde and retrograde orbits. This is also expected as radial orbits have lower absolute angular momentum and thus higher radial velocity variance.

$\bullet$ Moving to the energy space, we calculated the $\beta$ energy profile and compared that with the $\beta$ radial profile where we made another halo classification based on their similarities/differences. We explicitly showed that 65\% of halos which present very mild sensitivity on different selection criteria, in the real space, are among those for which the $\beta$ radial and energy profiles are very closely related. On the contrary, halos for which the $\beta$ radial/energy profile are rather different, are entirely parts of category $\mathcal{C}$. This again establishes that we should be very careful when drawing any conclusions from the $\beta$ analysis. 

$\bullet$ Furthermore, we proposed that one quantitative way to examine whether the observational results are somewhat robust is to also compute the $\beta$ energy profile and compare that with the $\beta$ radial profile. Based on our conclusion, it is likely that those with very similar profiles are more robust against altering the selection criteria than the ones with different profiles. 
However, as already indicated above, the current analysis and comparison are purely theoretical. From the observational perspective, we need to estimate the gravitational potential. The current paper is not meant to show that we identify the unbiased sample of stars using this comparison. Merely that for a chosen stellar sample, this comparison may indicate how much the chosen sample might be sensitive to different criteria.

$\bullet$  Ultimately, we overlaid our theoretical outcomes, with few different selection criteria, on top of some recent observational results and compared them with each other. Our comparison was made both at the level of isolated data points, in which each observation gives us only one point, as well as the extended observations in which the $\beta$ radial profile is reported in more than one single point for a particular observation. The latter one is more meaningful since we are not affected by the biases coming from changing the observations or the selection criteria. It is shown that there are reasonable agreements between the theoretical predictions and the observational results both for the isolated data points as well as the extended observations. It is seen for most of halos there are some data points that easily pass through the profile.

\section*{Data Availability}
Data directly related to this manuscript and its figures is available to share on reasonable requests from the corresponding author. The IllustrisTNG and TNG50 simulations are publicly available and accessible at \url{www.tng-project.org/data} \citep{2019ComAC...6....2N}.

\section*{acknowledgement}
It is a great pleasure to thank Sirio Belli, Sownak Bose, Shep Doeleman and Sandro Tacchella for the very insightful conversations. We are also grateful to the referee for very constructive report that improved the quality of this paper. Razieh Emami acknowledges the support by the Institute for Theory and Computation at the Center for Astrophysics. We thank the supercomputer facility at Harvard where most of the simulation work was done. 
MV acknowledges support through an MIT RSC award, a Kavli Research Investment Fund, NASA ATP grant NNX17AG29G, and NSF grants AST-1814053, AST-1814259 and AST-1909831. FM acknowledges support through the Program "Rita Levi Montalcini" of the Italian MIUR. The TNG50 simulation was realized with compute time granted by the Gauss center for Supercomputing (GCS) under GCS Large-Scale Projects GCS-DWAR on the GCS share of the supercomputer Hazel Hen at the High Performance Computing Center Stuttgart (HLRS).

\textit{Software:} matplotlib \citep{2007CSE.....9...90H}, numpy \citep{2011CSE....13b..22V}, scipy \citep{2007CSE.....9c..10O}, seaborn \citep{2020zndo...3629446W}, pandas \citep{2021zndo...5203279R}, h5py \citep{2016arXiv160804904D}.

\bibliography{main}{}

\begin{thebibliography}{}
\expandafter\ifx\csname natexlab\endcsname\relax\def\natexlab#1{#1}\fi
\providecommand{\url}[1]{\href{#1}{#1}}
\providecommand{\dodoi}[1]{doi:~\href{http://doi.org/#1}{\nolinkurl{#1}}}
\providecommand{\doeprint}[1]{\href{http://ascl.net/#1}{\nolinkurl{http://ascl.net/#1}}}
\providecommand{\doarXiv}[1]{\href{https://arxiv.org/abs/#1}{\nolinkurl{https://arxiv.org/abs/#1}}}

\bibitem[{{Abadi} {et~al.}(2006){Abadi}, {Navarro}, \&
  {Steinmetz}}]{2006MNRAS.365..747A}
{Abadi}, M.~G., {Navarro}, J.~F., \& {Steinmetz}, M. 2006, \mnras, 365, 747,
  \dodoi{10.1111/j.1365-2966.2005.09789.x}

\bibitem[{{Adams} {et~al.}(2012){Adams}, {Gebhardt}, {Blanc}, {Fabricius},
  {Hill}, {Murphy}, {van den Bosch}, \& {van de Ven}}]{2012ApJ...745...92A}
{Adams}, J.~J., {Gebhardt}, K., {Blanc}, G.~A., {et~al.} 2012, \apj, 745, 92,
  \dodoi{10.1088/0004-637X/745/1/92}

\bibitem[{{Adams} {et~al.}(2014){Adams}, {Simon}, {Fabricius}, {van den Bosch},
  {Barentine}, {Bender}, {Gebhardt}, {Hill}, {Murphy}, {Swaters}, {Thomas}, \&
  {van de Ven}}]{2014ApJ...789...63A}
{Adams}, J.~J., {Simon}, J.~D., {Fabricius}, M.~H., {et~al.} 2014, \apj, 789,
  63, \dodoi{10.1088/0004-637X/789/1/63}

\bibitem[{{Anders} {et~al.}(2017){Anders}, {Chiappini}, {Minchev}, {Miglio},
  {Montalb{\'a}n}, {Mosser}, {Rodrigues}, {Santiago}, {Baudin}, {Beers}, {da
  Costa}, {Garc{\'\i}a}, {Garc{\'\i}a-Hern{\'a}ndez}, {Holtzman}, {Maia},
  {Majewski}, {Mathur}, {Noels-Grotsch}, {Pan}, {Schneider}, {Schultheis},
  {Steinmetz}, {Valentini}, \& {Zamora}}]{2017A&A...600A..70A}
{Anders}, F., {Chiappini}, C., {Minchev}, I., {et~al.} 2017, \aap, 600, A70,
  \dodoi{10.1051/0004-6361/201629363}

\bibitem[{{Battaglia} {et~al.}(2008){Battaglia}, {Helmi}, {Tolstoy}, {Irwin},
  {Hill}, \& {Jablonka}}]{2008ApJ...681L..13B}
{Battaglia}, G., {Helmi}, A., {Tolstoy}, E., {et~al.} 2008, \apjl, 681, L13,
  \dodoi{10.1086/590179}

\bibitem[{{Beers} {et~al.}(2012){Beers}, {Carollo}, {Ivezi{\'c}}, {An},
  {Chiba}, {Norris}, {Freeman}, {Lee}, {Munn}, {Re Fiorentin}, {Sivarani},
  {Wilhelm}, {Yanny}, \& {York}}]{2012ApJ...746...34B}
{Beers}, T.~C., {Carollo}, D., {Ivezi{\'c}}, {\v{Z}}., {et~al.} 2012, \apj,
  746, 34, \dodoi{10.1088/0004-637X/746/1/34}

\bibitem[{{Binney}(1980)}]{1980MNRAS.190..873B}
{Binney}, J. 1980, \mnras, 190, 873, \dodoi{10.1093/mnras/190.4.873}

\bibitem[{{Bird} {et~al.}(2021){Bird}, {Loebman}, {Weinberg}, {Brooks},
  {Quinn}, \& {Christensen}}]{2021MNRAS.503.1815B}
{Bird}, J.~C., {Loebman}, S.~R., {Weinberg}, D.~H., {et~al.} 2021, \mnras, 503,
  1815, \dodoi{10.1093/mnras/stab289}

\bibitem[{{Bird} {et~al.}(2019){Bird}, {Xue}, {Liu}, {Shen}, {Flynn}, \&
  {Yang}}]{2019AJ....157..104B}
{Bird}, S.~A., {Xue}, X.-X., {Liu}, C., {et~al.} 2019, \aj, 157, 104,
  \dodoi{10.3847/1538-3881/aafd2e}

\bibitem[{{Bird} {et~al.}(2020){Bird}, {Xue}, {Liu}, {Shen}, {Flynn}, {Yang},
  {Zhao}, \& {Tian}}]{2020arXiv200505980B}
---. 2020, arXiv e-prints, arXiv:2005.05980.
\newblock \doarXiv{2005.05980}

\bibitem[{{Bond} {et~al.}(2010){Bond}, {Ivezi{\'c}}, {Sesar}, {Juri{\'c}},
  {Munn}, {Kowalski}, {Loebman}, {Ro{\v{s}}kar}, {Beers}, {Dalcanton},
  {Rockosi}, {Yanny}, {Newberg}, {Allende Prieto}, {Wilhelm}, {Lee},
  {Sivarani}, {Majewski}, {Norris}, {Bailer-Jones}, {Re Fiorentin}, {Schlegel},
  {Uomoto}, {Lupton}, {Knapp}, {Gunn}, {Covey}, {Allyn Smith}, {Miknaitis},
  {Doi}, {Tanaka}, {Fukugita}, {Kent}, {Finkbeiner}, {Quinn}, {Hawley},
  {Anderson}, {Kiuchi}, {Chen}, {Bushong}, {Sohi}, {Haggard}, {Kimball},
  {McGurk}, {Barentine}, {Brewington}, {Harvanek}, {Kleinman}, {Krzesinski},
  {Long}, {Nitta}, {Snedden}, {Lee}, {Pier}, {Harris}, {Brinkmann}, \&
  {Schneider}}]{2010ApJ...716....1B}
{Bond}, N.~A., {Ivezi{\'c}}, {\v{Z}}., {Sesar}, B., {et~al.} 2010, \apj, 716,
  1, \dodoi{10.1088/0004-637X/716/1/1}

\bibitem[{{Chiba} \& {Yoshii}(1998)}]{1998AJ....115..168C}
{Chiba}, M., \& {Yoshii}, Y. 1998, \aj, 115, 168, \dodoi{10.1086/300177}

\bibitem[{{Cinzano} \& {van der Marel}(1994)}]{1994MNRAS.270..325C}
{Cinzano}, P., \& {van der Marel}, R.~P. 1994, \mnras, 270, 325,
  \dodoi{10.1093/mnras/270.2.325}

\bibitem[{{Courteau} {et~al.}(2014){Courteau}, {Cappellari}, {de Jong},
  {Dutton}, {Emsellem}, {Hoekstra}, {Koopmans}, {Mamon}, {Maraston}, {Treu}, \&
  {Widrow}}]{2014RvMP...86...47C}
{Courteau}, S., {Cappellari}, M., {de Jong}, R.~S., {et~al.} 2014, Reviews of
  Modern Physics, 86, 47, \dodoi{10.1103/RevModPhys.86.47}

\bibitem[{{Cunningham} {et~al.}(2016){Cunningham}, {Deason}, {Guhathakurta},
  {Rockosi}, {van der Marel}, {Toloba}, {Gilbert}, {Sohn}, \&
  {Dorman}}]{2016ApJ...820...18C}
{Cunningham}, E.~C., {Deason}, A.~J., {Guhathakurta}, P., {et~al.} 2016, \apj,
  820, 18, \dodoi{10.3847/0004-637X/820/1/18}

\bibitem[{{Cunningham} {et~al.}(2019){Cunningham}, {Deason}, {Sanderson},
  {Sohn}, {Anderson}, {Guhathakurta}, {Rockosi}, {van der Marel}, {Loebman}, \&
  {Wetzel}}]{2019ApJ...879..120C}
{Cunningham}, E.~C., {Deason}, A.~J., {Sanderson}, R.~E., {et~al.} 2019, \apj,
  879, 120, \dodoi{10.3847/1538-4357/ab24cd}

\bibitem[{{de Buyl} {et~al.}(2016){de Buyl}, {Huang}, \&
  {Deprez}}]{2016arXiv160804904D}
{de Buyl}, P., {Huang}, M.-J., \& {Deprez}, L. 2016, arXiv e-prints,
  arXiv:1608.04904.
\newblock \doarXiv{1608.04904}

\bibitem[{{Deason} {et~al.}(2012){Deason}, {Belokurov}, {Evans}, \&
  {An}}]{2012MNRAS.424L..44D}
{Deason}, A.~J., {Belokurov}, V., {Evans}, N.~W., \& {An}, J. 2012, \mnras,
  424, L44, \dodoi{10.1111/j.1745-3933.2012.01283.x}

\bibitem[{{Dejonghe} \& {Merritt}(1992)}]{1992ApJ...391..531D}
{Dejonghe}, H., \& {Merritt}, D. 1992, \apj, 391, 531, \dodoi{10.1086/171368}

\bibitem[{{Dekel} {et~al.}(2005){Dekel}, {Stoehr}, {Mamon}, {Cox}, {Novak}, \&
  {Primack}}]{2005Natur.437..707D}
{Dekel}, A., {Stoehr}, F., {Mamon}, G.~A., {et~al.} 2005, \nat, 437, 707,
  \dodoi{10.1038/nature03970}

\bibitem[{{Diakogiannis} {et~al.}(2014{\natexlab{a}}){Diakogiannis}, {Lewis},
  \& {Ibata}}]{2014MNRAS.443..598D}
{Diakogiannis}, F.~I., {Lewis}, G.~F., \& {Ibata}, R.~A. 2014{\natexlab{a}},
  \mnras, 443, 598, \dodoi{10.1093/mnras/stu1153}

\bibitem[{{Diakogiannis} {et~al.}(2014{\natexlab{b}}){Diakogiannis}, {Lewis},
  \& {Ibata}}]{2014MNRAS.443..610D}
---. 2014{\natexlab{b}}, \mnras, 443, 610, \dodoi{10.1093/mnras/stu1154}

\bibitem[{{Diemand} {et~al.}(2005){Diemand}, {Madau}, \&
  {Moore}}]{2005MNRAS.364..367D}
{Diemand}, J., {Madau}, P., \& {Moore}, B. 2005, \mnras, 364, 367,
  \dodoi{10.1111/j.1365-2966.2005.09604.x}

\bibitem[{{Dietz} {et~al.}(2020){Dietz}, {Yoon}, {Beers}, \&
  {Placco}}]{2020ApJ...894...34D}
{Dietz}, S.~E., {Yoon}, J., {Beers}, T.~C., \& {Placco}, V.~M. 2020, \apj, 894,
  34, \dodoi{10.3847/1538-4357/ab7fa4}

\bibitem[{{El-Badry} {et~al.}(2017){El-Badry}, {Wetzel}, {Geha}, {Quataert},
  {Hopkins}, {Kere{\v{s}}}, {Chan}, \&
  {Faucher-Gigu{\`e}re}}]{2017ApJ...835..193E}
{El-Badry}, K., {Wetzel}, A.~R., {Geha}, M., {et~al.} 2017, \apj, 835, 193,
  \dodoi{10.3847/1538-4357/835/2/193}

\bibitem[{{Emami} {et~al.}(2020){Emami}, {Hernquist}, {Alcock}, {Genel},
  {Bose}, {Weinberger}, {Vogelsberger}, {Shen}, {Speagle}, {Marinacci},
  {Forbes}, \& {Torrey}}]{2020arXiv201212284E}
{Emami}, R., {Hernquist}, L., {Alcock}, C., {et~al.} 2020, arXiv e-prints,
  arXiv:2012.12284.
\newblock \doarXiv{2012.12284}

\bibitem[{{Emami} {et~al.}(2021{\natexlab{a}}){Emami}, {Genel}, {Hernquist},
  {Alcock}, {Bose}, {Weinberger}, {Vogelsberger}, {Marinacci}, {Loeb},
  {Torrey}, \& {Forbes}}]{2021ApJ...913...36E}
{Emami}, R., {Genel}, S., {Hernquist}, L., {et~al.} 2021{\natexlab{a}}, \apj,
  913, 36, \dodoi{10.3847/1538-4357/abf147}

\bibitem[{{Emami} {et~al.}(2021{\natexlab{b}}){Emami}, {Hernquist}, {Alcock},
  {Genel}, {Bose}, {Weinberger}, {Vogelsberger}, {Shen}, {Speagle},
  {Marinacci}, {Forbes}, \& {Torrey}}]{2021AAS...23832407E}
{Emami}, R., {Hernquist}, L., {Alcock}, C., {et~al.} 2021{\natexlab{b}}, in
  American Astronomical Society Meeting Abstracts, Vol.~53, American
  Astronomical Society Meeting Abstracts, 324.07

\bibitem[{{Frankel} {et~al.}(2019){Frankel}, {Sanders}, {Rix}, {Ting}, \&
  {Ness}}]{2019ApJ...884...99F}
{Frankel}, N., {Sanders}, J., {Rix}, H.-W., {Ting}, Y.-S., \& {Ness}, M. 2019,
  \apj, 884, 99, \dodoi{10.3847/1538-4357/ab4254}

\bibitem[{{Gaia Collaboration} {et~al.}(2018){Gaia Collaboration}, {Helmi},
  {van Leeuwen}, {McMillan}, {Massari}, {Antoja}, {Robin}, {Lindegren},
  {Bastian}, {Arenou}, {Babusiaux}, {Biermann}, {Breddels}, {Hobbs}, {Jordi},
  {Pancino}, {Reyl{\'e}}, {Veljanoski}, {Brown}, {Vallenari}, {Prusti}, {de
  Bruijne}, {Bailer-Jones}, {Evans}, {Eyer}, {Jansen}, {Klioner}, {Lammers},
  {Luri}, {Mignard}, {Panem}, {Pourbaix}, {Randich}, {Sartoretti}, {Siddiqui},
  {Soubiran}, {Walton}, {Cropper}, {Drimmel}, {Katz}, {Lattanzi}, {Bakker},
  {Cacciari}, {Casta{\~n}eda}, {Chaoul}, {Cheek}, {De Angeli}, {Fabricius},
  {Guerra}, {Holl}, {Masana}, {Messineo}, {Mowlavi}, {Nienartowicz}, {Panuzzo},
  {Portell}, {Riello}, {Seabroke}, {Tanga}, {Th{\'e}venin}, {Gracia-Abril},
  {Comoretto}, {Garcia-Reinaldos}, {Teyssier}, {Altmann}, {Andrae}, {Audard},
  {Bellas-Velidis}, {Benson}, {Berthier}, {Blomme}, {Burgess}, {Busso},
  {Carry}, {Cellino}, {Clementini}, {Clotet}, {Creevey}, {Davidson}, {De
  Ridder}, {Delchambre}, {Dell'Oro}, {Ducourant},
  {Fern{\'a}ndez-Hern{\'a}ndez}, {Fouesneau}, {Fr{\'e}mat}, {Galluccio},
  {Garc{\'\i}a-Torres}, {Gonz{\'a}lez-N{\'u}{\~n}ez}, {Gonz{\'a}lez-Vidal},
  {Gosset}, {Guy}, {Halbwachs}, {Hambly}, {Harrison}, {Hern{\'a}ndez},
  {Hestroffer}, {Hodgkin}, {Hutton}, {Jasniewicz}, {Jean-Antoine-Piccolo},
  {Jordan}, {Korn}, {Krone-Martins}, {Lanzafame}, {Lebzelter}, {L{\"o}ffler},
  {Manteiga}, {Marrese}, {Mart{\'\i}n-Fleitas}, {Moitinho}, {Mora}, {Muinonen},
  {Osinde}, {Pauwels}, {Petit}, {Recio-Blanco}, {Richards}, {Rimoldini},
  {Sarro}, {Siopis}, {Smith}, {Sozzetti}, {S{\"u}veges}, {Torra}, {van Reeven},
  {Abbas}, {Abreu Aramburu}, {Accart}, {Aerts}, {Altavilla}, {{\'A}lvarez},
  {Alvarez}, {Alves}, {Anderson}, {Andrei}, {Anglada Varela}, {Antiche},
  {Arcay}, {Astraatmadja}, {Bach}, {Baker}, {Balaguer-N{\'u}{\~n}ez}, {Balm},
  {Barache}, {Barata}, {Barbato}, {Barblan}, {Barklem}, {Barrado}, {Barros},
  {Barstow}, {Bartholom{\'e} Mu{\~n}oz}, {Bassilana}, {Becciani}, {Bellazzini},
  {Berihuete}, {Bertone}, {Bianchi}, {Bienaym{\'e}}, {Blanco-Cuaresma}, {Boch},
  {Boeche}, {Bombrun}, {Borrachero}, {Bossini}, {Bouquillon}, {Bourda},
  {Bragaglia}, {Bramante}, {Bressan}, {Brouillet}, {Br{\"u}semeister},
  {Brugaletta}, {Bucciarelli}, {Burlacu}, {Busonero}, {Butkevich}, {Buzzi},
  {Caffau}, {Cancelliere}, {Cannizzaro}, {Cantat-Gaudin}, {Carballo},
  {Carlucci}, {Carrasco}, {Casamiquela}, {Castellani}, {Castro-Ginard},
  {Charlot}, {Chemin}, {Chiavassa}, {Cocozza}, {Costigan}, {Cowell}, {Crifo},
  {Crosta}, {Crowley}, {Cuypers}, {Dafonte}, {Damerdji}, {Dapergolas}, {David},
  {David}, {de Laverny}, {De Luise}, {De March}, {de Martino}, {de Souza}, {de
  Torres}, {Debosscher}, {del Pozo}, {Delbo}, {Delgado}, {Delgado}, {Di
  Matteo}, {Diakite}, {Diener}, {Distefano}, {Dolding}, {Drazinos},
  {Dur{\'a}n}, {Edvardsson}, {Enke}, {Eriksson}, {Esquej}, {Eynard Bontemps},
  {Fabre}, {Fabrizio}, {Faigler}, {Falc{\~a}o}, {Farr{\`a}s Casas}, {Federici},
  {Fedorets}, {Fernique}, {Figueras}, {Filippi}, {Findeisen}, {Fonti},
  {Fraile}, {Fraser}, {Fr{\'e}zouls}, {Gai}, {Galleti}, {Garabato},
  {Garc{\'\i}a-Sedano}, {Garofalo}, {Garralda}, {Gavel}, {Gavras}, {Gerssen},
  {Geyer}, {Giacobbe}, {Gilmore}, {Girona}, {Giuffrida}, {Glass}, {Gomes},
  {Granvik}, {Gueguen}, {Guerrier}, {Guiraud}, {Guti{\'e}rrez-S{\'a}nchez},
  {Hofmann}, {Holland}, {Huckle}, {Hypki}, {Icardi}, {Jan{\ss}en}, {Jevardat de
  Fombelle}, {Jonker}, {Juh{\'a}sz}, {Julbe}, {Karampelas}, {Kewley}, {Klar},
  {Kochoska}, {Kohley}, {Kolenberg}, {Kontizas}, {Kontizas}, {Koposov},
  {Kordopatis}, {Kostrzewa-Rutkowska}, {Koubsky}, {Lambert}, {Lanza}, {Lasne},
  {Lavigne}, {Le Fustec}, {Le Poncin-Lafitte}, {Lebreton}, {Leccia}, {Leclerc},
  {Lecoeur-Taibi}, {Lenhardt}, {Leroux}, {Liao}, {Licata}, {Lindstr{\o}m},
  {Lister}, {Livanou}, {Lobel}, {L{\'o}pez}, {Managau}, {Mann}, {Mantelet},
  {Marchal}, {Marchant}, {Marconi}, {Marinoni}, {Marschalk{\'o}}, {Marshall},
  {Martino}, {Marton}, {Mary}, {Matijevi{\v{c}}}, {Mazeh}, {Messina},
  {Michalik}, {Millar}, {Molina}, {Molinaro}, {Moln{\'a}r}, {Montegriffo},
  {Mor}, {Morbidelli}, {Morel}, {Morris}, {Mulone}, {Muraveva}, {Musella},
  {Nelemans}, {Nicastro}, {Noval}, {O'Mullane}, {Ord{\'e}novic},
  {Ord{\'o}{\~n}ez-Blanco}, {Osborne}, {Pagani}, {Pagano}, {Pailler},
  {Palacin}, {Palaversa}, {Panahi}, {Pawlak}, {Piersimoni}, {Pineau}, {Plachy},
  {Plum}, {Poggio}, {Poujoulet}, {Pr{\v{s}}a}, {Pulone}, {Racero}, {Ragaini},
  {Rambaux}, {Ramos-Lerate}, {Regibo}, {Riclet}, {Ripepi}, {Riva}, {Rivard},
  {Rixon}, {Roegiers}, {Roelens}, {Romero-G{\'o}mez}, {Rowell}, {Royer},
  {Ruiz-Dern}, {Sadowski}, {Sagrist{\`a} Sell{\'e}s}, {Sahlmann}, {Salgado},
  {Salguero}, {Sanna}, {Santana-Ros}, {Sarasso}, {Savietto}, {Schultheis},
  {Sciacca}, {Segol}, {Segovia}, {S{\'e}gransan}, {Shih}, {Siltala}, {Silva},
  {Smart}, {Smith}, {Solano}, {Solitro}, {Sordo}, {Soria Nieto}, {Souchay},
  {Spagna}, {Spoto}, {Stampa}, {Steele}, {Steidelm{\"u}ller}, {Stephenson},
  {Stoev}, {Suess}, {Surdej}, {Szabados}, {Szegedi-Elek}, {Tapiador}, {Taris},
  {Tauran}, {Taylor}, {Teixeira}, {Terrett}, {Teyssandier}, {Thuillot},
  {Titarenko}, {Torra Clotet}, {Turon}, {Ulla}, {Utrilla}, {Uzzi}, {Vaillant},
  {Valentini}, {Valette}, {van Elteren}, {Van Hemelryck}, {van Leeuwen},
  {Vaschetto}, {Vecchiato}, {Viala}, {Vicente}, {Vogt}, {von Essen}, {Voss},
  {Votruba}, {Voutsinas}, {Walmsley}, {Weiler}, {Wertz}, {Wevems},
  {Wyrzykowski}, {Yoldas}, {{\v{Z}}erjal}, {Ziaeepour}, {Zorec}, {Zschocke},
  {Zucker}, {Zurbach}, \& {Zwitter}}]{2018A&A...616A..12G}
{Gaia Collaboration}, {Helmi}, A., {van Leeuwen}, F., {et~al.} 2018, \aap, 616,
  A12, \dodoi{10.1051/0004-6361/201832698}

\bibitem[{{Gallazzi} {et~al.}(2005){Gallazzi}, {Charlot}, {Brinchmann},
  {White}, \& {Tremonti}}]{2005MNRAS.362...41G}
{Gallazzi}, A., {Charlot}, S., {Brinchmann}, J., {White}, S. D.~M., \&
  {Tremonti}, C.~A. 2005, \mnras, 362, 41,
  \dodoi{10.1111/j.1365-2966.2005.09321.x}

\bibitem[{{Gilmore} {et~al.}(2007){Gilmore}, {Wilkinson}, {Wyse}, {Kleyna},
  {Koch}, {Evans}, \& {Grebel}}]{2007ApJ...663..948G}
{Gilmore}, G., {Wilkinson}, M.~I., {Wyse}, R. F.~G., {et~al.} 2007, \apj, 663,
  948, \dodoi{10.1086/518025}

\bibitem[{{Grand} {et~al.}(2018){Grand}, {Helly}, {Fattahi}, {Cautun}, {Cole},
  {Cooper}, {Deason}, {Frenk}, {G{\'o}mez}, {Hunt}, {Marinacci}, {Pakmor},
  {Simpson}, {Springel}, \& {Xu}}]{2018MNRAS.481.1726G}
{Grand}, R. J.~J., {Helly}, J., {Fattahi}, A., {et~al.} 2018, \mnras, 481,
  1726, \dodoi{10.1093/mnras/sty2403}

\bibitem[{{Hani} {et~al.}(2019){Hani}, {Ellison}, {Sparre}, {Grand}, {Pakmor},
  {Gomez}, \& {Springel}}]{2019MNRAS.488..135H}
{Hani}, M.~H., {Ellison}, S.~L., {Sparre}, M., {et~al.} 2019, \mnras, 488, 135,
  \dodoi{10.1093/mnras/stz1708}

\bibitem[{{Hattori} {et~al.}(2017){Hattori}, {Valluri}, {Loebman}, \&
  {Bell}}]{2017ApJ...841...91H}
{Hattori}, K., {Valluri}, M., {Loebman}, S.~R., \& {Bell}, E.~F. 2017, \apj,
  841, 91, \dodoi{10.3847/1538-4357/aa71aa}

\bibitem[{{Hattori} {et~al.}(2013){Hattori}, {Yoshii}, {Beers}, {Carollo}, \&
  {Lee}}]{2013ApJ...763L..17H}
{Hattori}, K., {Yoshii}, Y., {Beers}, T.~C., {Carollo}, D., \& {Lee}, Y.~S.
  2013, \apjl, 763, L17, \dodoi{10.1088/2041-8205/763/1/L17}

\bibitem[{{Huang} {et~al.}(2015){Huang}, {Liu}, {Zhang}, {Yuan}, {Xiang},
  {Chen}, {Ren}, {Sun}, {Wang}, {Zhang}, {Hou}, {Wang}, \&
  {Yang}}]{2015RAA....15.1240H}
{Huang}, Y., {Liu}, X.-W., {Zhang}, H.-W., {et~al.} 2015, Research in Astronomy
  and Astrophysics, 15, 1240, \dodoi{10.1088/1674-4527/15/8/010}

\bibitem[{{Hunter}(2007)}]{2007CSE.....9...90H}
{Hunter}, J.~D. 2007, Computing in Science and Engineering, 9, 90,
  \dodoi{10.1109/MCSE.2007.55}

\bibitem[{{Jeans}(1915)}]{1915MNRAS..76...70J}
{Jeans}, J.~H. 1915, \mnras, 76, 70, \dodoi{10.1093/mnras/76.2.70}

\bibitem[{{Johnson} {et~al.}(2021){Johnson}, {Weinberg}, {Vincenzo}, {Bird},
  {Loebman}, {Brooks}, {Quinn}, {Christensen}, \&
  {Griffith}}]{2021arXiv210309838J}
{Johnson}, J.~W., {Weinberg}, D.~H., {Vincenzo}, F., {et~al.} 2021, arXiv
  e-prints, arXiv:2103.09838.
\newblock \doarXiv{2103.09838}

\bibitem[{{Kafle} {et~al.}(2012){Kafle}, {Sharma}, {Lewis}, \&
  {Bland-Hawthorn}}]{2012ApJ...761...98K}
{Kafle}, P.~R., {Sharma}, S., {Lewis}, G.~F., \& {Bland-Hawthorn}, J. 2012,
  \apj, 761, 98, \dodoi{10.1088/0004-637X/761/2/98}

\bibitem[{{Kewley} \& {Ellison}(2008)}]{2008ApJ...681.1183K}
{Kewley}, L.~J., \& {Ellison}, S.~L. 2008, \apj, 681, 1183,
  \dodoi{10.1086/587500}

\bibitem[{{King} {et~al.}(2015){King}, {Brown}, {Geller}, \&
  {Kenyon}}]{2015ApJ...813...89K}
{King}, Charles, I., {Brown}, W.~R., {Geller}, M.~J., \& {Kenyon}, S.~J. 2015,
  \apj, 813, 89, \dodoi{10.1088/0004-637X/813/2/89}

\bibitem[{{Kleyna} {et~al.}(2001){Kleyna}, {Wilkinson}, {Evans}, \&
  {Gilmore}}]{2001ApJ...563L.115K}
{Kleyna}, J.~T., {Wilkinson}, M.~I., {Evans}, N.~W., \& {Gilmore}, G. 2001,
  \apjl, 563, L115, \dodoi{10.1086/338603}

\bibitem[{{Koch} {et~al.}(2007){Koch}, {Kleyna}, {Wilkinson}, {Grebel},
  {Gilmore}, {Evans}, {Wyse}, \& {Harbeck}}]{2007AJ....134..566K}
{Koch}, A., {Kleyna}, J.~T., {Wilkinson}, M.~I., {et~al.} 2007, \aj, 134, 566,
  \dodoi{10.1086/519380}

\bibitem[{{Krajnovi{\'c}} {et~al.}(2005){Krajnovi{\'c}}, {Cappellari},
  {Emsellem}, {McDermid}, \& {de Zeeuw}}]{2005MNRAS.357.1113K}
{Krajnovi{\'c}}, D., {Cappellari}, M., {Emsellem}, E., {McDermid}, R.~M., \&
  {de Zeeuw}, P.~T. 2005, \mnras, 357, 1113,
  \dodoi{10.1111/j.1365-2966.2005.08715.x}

\bibitem[{{Lancaster} {et~al.}(2019){Lancaster}, {Koposov}, {Belokurov},
  {Evans}, \& {Deason}}]{2019MNRAS.486..378L}
{Lancaster}, L., {Koposov}, S.~E., {Belokurov}, V., {Evans}, N.~W., \&
  {Deason}, A.~J. 2019, \mnras, 486, 378, \dodoi{10.1093/mnras/stz853}

\bibitem[{{Loebman} {et~al.}(2018){Loebman}, {Valluri}, {Hattori},
  {Debattista}, {Bell}, {Stinson}, {Christensen}, {Brooks}, {Quinn}, \&
  {Governato}}]{2018ApJ...853..196L}
{Loebman}, S.~R., {Valluri}, M., {Hattori}, K., {et~al.} 2018, \apj, 853, 196,
  \dodoi{10.3847/1538-4357/aaa0d6}

\bibitem[{{{\L}okas}(2009)}]{2009MNRAS.394L.102L}
{{\L}okas}, E.~L. 2009, \mnras, 394, L102,
  \dodoi{10.1111/j.1745-3933.2009.00620.x}

\bibitem[{{{\L}okas} {et~al.}(2005){{\L}okas}, {Mamon}, \&
  {Prada}}]{2005MNRAS.363..918L}
{{\L}okas}, E.~L., {Mamon}, G.~A., \& {Prada}, F. 2005, \mnras, 363, 918,
  \dodoi{10.1111/j.1365-2966.2005.09497.x}

\bibitem[{{Mackereth} {et~al.}(2019){Mackereth}, {Schiavon}, {Pfeffer},
  {Hayes}, {Bovy}, {Anguiano}, {Allende Prieto}, {Hasselquist}, {Holtzman},
  {Johnson}, {Majewski}, {O'Connell}, {Shetrone}, {Tissera}, \&
  {Fern{\'a}ndez-Trincado}}]{2019MNRAS.482.3426M}
{Mackereth}, J.~T., {Schiavon}, R.~P., {Pfeffer}, J., {et~al.} 2019, \mnras,
  482, 3426, \dodoi{10.1093/mnras/sty2955}

\bibitem[{{Mamon} {et~al.}(2013){Mamon}, {Biviano}, \&
  {Bou{\'e}}}]{2013MNRAS.429.3079M}
{Mamon}, G.~A., {Biviano}, A., \& {Bou{\'e}}, G. 2013, \mnras, 429, 3079,
  \dodoi{10.1093/mnras/sts565}

\bibitem[{{Mamon} \& {{\L}okas}(2005)}]{2005MNRAS.363..705M}
{Mamon}, G.~A., \& {{\L}okas}, E.~L. 2005, \mnras, 363, 705,
  \dodoi{10.1111/j.1365-2966.2005.09400.x}

\bibitem[{{Mashchenko}(2015)}]{2015arXiv150408273M}
{Mashchenko}, S. 2015, arXiv e-prints, arXiv:1504.08273.
\newblock \doarXiv{1504.08273}

\bibitem[{{Merritt}(1985)}]{1985AJ.....90.1027M}
{Merritt}, D. 1985, \aj, 90, 1027, \dodoi{10.1086/113810}

\bibitem[{{Merritt}(2015)}]{2015ApJ...814...57M}
---. 2015, \apj, 814, 57, \dodoi{10.1088/0004-637X/814/1/57}

\bibitem[{{Monachesi} {et~al.}(2016){Monachesi}, {G{\'o}mez}, {Grand},
  {Kauffmann}, {Marinacci}, {Pakmor}, {Springel}, \&
  {Frenk}}]{2016MNRAS.459L..46M}
{Monachesi}, A., {G{\'o}mez}, F.~A., {Grand}, R. J.~J., {et~al.} 2016, \mnras,
  459, L46, \dodoi{10.1093/mnrasl/slw052}

\bibitem[{{Monachesi} {et~al.}(2019){Monachesi}, {G{\'o}mez}, {Grand},
  {Simpson}, {Kauffmann}, {Bustamante}, {Marinacci}, {Pakmor}, {Springel},
  {Frenk}, {White}, \& {Tissera}}]{2019MNRAS.485.2589M}
---. 2019, \mnras, 485, 2589, \dodoi{10.1093/mnras/stz538}

\bibitem[{{Morrison} {et~al.}(1990){Morrison}, {Flynn}, \&
  {Freeman}}]{1990AJ....100.1191M}
{Morrison}, H.~L., {Flynn}, C., \& {Freeman}, K.~C. 1990, \aj, 100, 1191,
  \dodoi{10.1086/115587}

\bibitem[{{Naidu} {et~al.}(2020){Naidu}, {Conroy}, {Bonaca}, {Johnson}, {Ting},
  {Caldwell}, {Zaritsky}, \& {Cargile}}]{2020ApJ...901...48N}
{Naidu}, R.~P., {Conroy}, C., {Bonaca}, A., {et~al.} 2020, \apj, 901, 48,
  \dodoi{10.3847/1538-4357/abaef4}

\bibitem[{{Nelson} {et~al.}(2019{\natexlab{a}}){Nelson}, {Pillepich},
  {Springel}, {Pakmor}, {Weinberger}, {Genel}, {Torrey}, {Vogelsberger},
  {Marinacci}, \& {Hernquist}}]{2019MNRAS.490.3234N}
{Nelson}, D., {Pillepich}, A., {Springel}, V., {et~al.} 2019{\natexlab{a}},
  \mnras, 490, 3234, \dodoi{10.1093/mnras/stz2306}

\bibitem[{{Nelson} {et~al.}(2019{\natexlab{b}}){Nelson}, {Springel},
  {Pillepich}, {Rodriguez-Gomez}, {Torrey}, {Genel}, {Vogelsberger}, {Pakmor},
  {Marinacci}, {Weinberger}, {Kelley}, {Lovell}, {Diemer}, \&
  {Hernquist}}]{2019ComAC...6....2N}
{Nelson}, D., {Springel}, V., {Pillepich}, A., {et~al.} 2019{\natexlab{b}},
  Computational Astrophysics and Cosmology, 6, 2,
  \dodoi{10.1186/s40668-019-0028-x}

\bibitem[{{Oliphant}(2007)}]{2007CSE.....9c..10O}
{Oliphant}, T.~E. 2007, Computing in Science and Engineering, 9, 10,
  \dodoi{10.1109/MCSE.2007.58}

\bibitem[{{Pillepich} {et~al.}(2019){Pillepich}, {Nelson}, {Springel},
  {Pakmor}, {Torrey}, {Weinberger}, {Vogelsberger}, {Marinacci}, {Genel}, {van
  der Wel}, \& {Hernquist}}]{2019MNRAS.490.3196P}
{Pillepich}, A., {Nelson}, D., {Springel}, V., {et~al.} 2019, \mnras, 490,
  3196, \dodoi{10.1093/mnras/stz2338}

\bibitem[{{Rashkov} {et~al.}(2013){Rashkov}, {Pillepich}, {Deason}, {Madau},
  {Rockosi}, {Guedes}, \& {Mayer}}]{2013ApJ...773L..32R}
{Rashkov}, V., {Pillepich}, A., {Deason}, A.~J., {et~al.} 2013, \apjl, 773,
  L32, \dodoi{10.1088/2041-8205/773/2/L32}

\bibitem[{{Reback} {et~al.}(2021){Reback}, {jbrockmendel}, {McKinney}, {Van den
  Bossche}, {Augspurger}, {Cloud}, {Hawkins}, {gfyoung}, {Sinhrks}, {Roeschke},
  {Klein}, {Petersen}, {Tratner}, {She}, {Ayd}, {Hoefler}, {Naveh}, {Garcia},
  {Schendel}, {Hayden}, {Saxton}, {Shadrach}, {Gorelli}, {Li}, {Jancauskas},
  {attack68}, {McMaster}, {Battiston}, {Seabold}, \&
  {Dong}}]{2021zndo...5203279R}
{Reback}, J., {jbrockmendel}, {McKinney}, W., {et~al.} 2021,
  {pandas-dev/pandas: Pandas 1.3.2}, v1.3.2,  Zenodo,
  \dodoi{10.5281/zenodo.5203279}

\bibitem[{{Sales} {et~al.}(2007){Sales}, {Navarro}, {Abadi}, \&
  {Steinmetz}}]{2007MNRAS.379.1464S}
{Sales}, L.~V., {Navarro}, J.~F., {Abadi}, M.~G., \& {Steinmetz}, M. 2007,
  \mnras, 379, 1464, \dodoi{10.1111/j.1365-2966.2007.12024.x}

\bibitem[{{Santucci} {et~al.}(2020){Santucci}, {Brough}, {Scott}, {Montes},
  {Owers}, {van Sande}, {Bland-Hawthorn}, {Bryant}, {Croom}, {Ferreras},
  {Lawrence}, {L{\'o}pez-S{\'a}nchez}, \& {Richards}}]{2020ApJ...896...75S}
{Santucci}, G., {Brough}, S., {Scott}, N., {et~al.} 2020, \apj, 896, 75,
  \dodoi{10.3847/1538-4357/ab92a9}

\bibitem[{{Schuster} {et~al.}(2012){Schuster}, {Moreno}, {Nissen}, \&
  {Pichardo}}]{2012A&A...538A..21S}
{Schuster}, W.~J., {Moreno}, E., {Nissen}, P.~E., \& {Pichardo}, B. 2012, \aap,
  538, A21, \dodoi{10.1051/0004-6361/201118035}

\bibitem[{{Sirko} {et~al.}(2004){Sirko}, {Goodman}, {Knapp}, {Brinkmann},
  {Ivezi{\'c}}, {Knerr}, {Schlegel}, {Schneider}, \&
  {York}}]{2004AJ....127..914S}
{Sirko}, E., {Goodman}, J., {Knapp}, G.~R., {et~al.} 2004, \aj, 127, 914,
  \dodoi{10.1086/381486}

\bibitem[{{Smith} {et~al.}(2009){Smith}, {Evans}, {Belokurov}, {Hewett},
  {Bramich}, {Gilmore}, {Irwin}, {Vidrih}, \& {Zucker}}]{2009MNRAS.399.1223S}
{Smith}, M.~C., {Evans}, N.~W., {Belokurov}, V., {et~al.} 2009, \mnras, 399,
  1223, \dodoi{10.1111/j.1365-2966.2009.15391.x}

\bibitem[{{Sohn} {et~al.}(2012){Sohn}, {Anderson}, \& {van der
  Marel}}]{2012ApJ...753....7S}
{Sohn}, S.~T., {Anderson}, J., \& {van der Marel}, R.~P. 2012, \apj, 753, 7,
  \dodoi{10.1088/0004-637X/753/1/7}

\bibitem[{{Sohn} {et~al.}(2018){Sohn}, {Watkins}, {Fardal}, {van der Marel},
  {Deason}, {Besla}, \& {Bellini}}]{2018ApJ...862...52S}
{Sohn}, S.~T., {Watkins}, L.~L., {Fardal}, M.~A., {et~al.} 2018, \apj, 862, 52,
  \dodoi{10.3847/1538-4357/aacd0b}

\bibitem[{{Sommer-Larsen} {et~al.}(1994){Sommer-Larsen}, {Flynn}, \&
  {Christensen}}]{1994MNRAS.271...94S}
{Sommer-Larsen}, J., {Flynn}, C., \& {Christensen}, P.~R. 1994, \mnras, 271,
  94, \dodoi{10.1093/mnras/271.1.94}

\bibitem[{{Stinson} {et~al.}(2013){Stinson}, {Brook}, {Macci{\`o}}, {Wadsley},
  {Quinn}, \& {Couchman}}]{2013MNRAS.428..129S}
{Stinson}, G.~S., {Brook}, C., {Macci{\`o}}, A.~V., {et~al.} 2013, \mnras, 428,
  129, \dodoi{10.1093/mnras/sts028}

\bibitem[{{Strigari} {et~al.}(2007){Strigari}, {Bullock}, {Kaplinghat},
  {Diemand}, {Kuhlen}, \& {Madau}}]{2007ApJ...669..676S}
{Strigari}, L.~E., {Bullock}, J.~S., {Kaplinghat}, M., {et~al.} 2007, \apj,
  669, 676, \dodoi{10.1086/521914}

\bibitem[{{Thom} {et~al.}(2005){Thom}, {Flynn}, {Bessell}, {H{\"a}nninen},
  {Beers}, {Christlieb}, {James}, {Holmberg}, \&
  {Gibson}}]{2005MNRAS.360..354T}
{Thom}, C., {Flynn}, C., {Bessell}, M.~S., {et~al.} 2005, \mnras, 360, 354,
  \dodoi{10.1111/j.1365-2966.2005.09038.x}

\bibitem[{{Tully} \& {Fisher}(1977)}]{1977A&A....54..661T}
{Tully}, R.~B., \& {Fisher}, J.~R. 1977, \aap, 500, 105

\bibitem[{{van der Walt} {et~al.}(2011){van der Walt}, {Colbert}, \&
  {Varoquaux}}]{2011CSE....13b..22V}
{van der Walt}, S., {Colbert}, S.~C., \& {Varoquaux}, G. 2011, Computing in
  Science and Engineering, 13, 22, \dodoi{10.1109/MCSE.2011.37}

\bibitem[{{Walker} {et~al.}(2007){Walker}, {Mateo}, {Olszewski}, {Gnedin},
  {Wang}, {Sen}, \& {Woodroofe}}]{2007ApJ...667L..53W}
{Walker}, M.~G., {Mateo}, M., {Olszewski}, E.~W., {et~al.} 2007, \apjl, 667,
  L53, \dodoi{10.1086/521998}

\bibitem[{{Walker} {et~al.}(2009){Walker}, {Mateo}, {Olszewski},
  {Pe{\~n}arrubia}, {Evans}, \& {Gilmore}}]{2009ApJ...704.1274W}
---. 2009, \apj, 704, 1274, \dodoi{10.1088/0004-637X/704/2/1274}

\bibitem[{{Waskom} {et~al.}(2020){Waskom}, {Botvinnik}, {Ostblom}, {Lukauskas},
  {Hobson}, {MaozGelbart}, {Gemperline}, {Augspurger}, {Halchenko}, {Cole},
  {Warmenhoven}, {De Ruiter}, {Pye}, {Hoyer}, {Vanderplas}, {Villalba},
  {Kunter}, {Quintero}, {Bachant}, {Martin}, {Meyer}, {Swain}, {Miles},
  {Brunner}, {O'Kane}, {Yarkoni}, {Williams}, \& {Evans}}]{2020zndo...3629446W}
{Waskom}, M., {Botvinnik}, O., {Ostblom}, J., {et~al.} 2020, {mwaskom/seaborn:
  v0.10.0 (January 2020)}, v0.10.0,  Zenodo, \dodoi{10.5281/zenodo.3629446}

\bibitem[{{Watkins} {et~al.}(2019){Watkins}, {van der Marel}, {Sohn}, \&
  {Evans}}]{2019ApJ...873..118W}
{Watkins}, L.~L., {van der Marel}, R.~P., {Sohn}, S.~T., \& {Evans}, N.~W.
  2019, \apj, 873, 118, \dodoi{10.3847/1538-4357/ab089f}

\bibitem[{{Wilkinson} {et~al.}(2004){Wilkinson}, {Kleyna}, {Evans}, {Gilmore},
  {Irwin}, \& {Grebel}}]{2004ApJ...611L..21W}
{Wilkinson}, M.~I., {Kleyna}, J.~T., {Evans}, N.~W., {et~al.} 2004, \apjl, 611,
  L21, \dodoi{10.1086/423619}

\bibitem[{{Wolf} {et~al.}(2010){Wolf}, {Martinez}, {Bullock}, {Kaplinghat},
  {Geha}, {Mu{\~n}oz}, {Simon}, \& {Avedo}}]{2010MNRAS.406.1220W}
{Wolf}, J., {Martinez}, G.~D., {Bullock}, J.~S., {et~al.} 2010, \mnras, 406,
  1220, \dodoi{10.1111/j.1365-2966.2010.16753.x}

\bibitem[{{Zuo} {et~al.}(2017){Zuo}, {Du}, {Jing}, {Gu}, {Newberg}, {Wu}, {Ma},
  \& {Zhou}}]{2017ApJ...841...59Z}
{Zuo}, W., {Du}, C., {Jing}, Y., {et~al.} 2017, \apj, 841, 59,
  \dodoi{10.3847/1538-4357/aa70e6}

\end{thebibliography}
\bibliographystyle{aasjournal}

\end{document}